\documentclass[twocolumn]{aastex62}
\pdfoutput=1
\usepackage{graphicx}
\usepackage[caption=false]{subfig}
\usepackage{amsmath}
\newcommand{\Rvir}{$R_{500}$\,}

\graphicspath{{./}{figures/}}

\shorttitle{Groups and Clusters near the NEP}
\shortauthors{Pratt and Bregman}

\begin{document}

\title{SZ Scaling Relations of Galaxy Groups and Clusters Near the North Ecliptic Pole}

\author{Cameron T. Pratt}
\author{Joel N. Bregman}
\affil{Department of Astronomy, University of Michigan, Ann Arbor, MI 48104}



\begin{abstract}

SZ scaling relations have been used to test the self-similar prediction for massive galaxy clusters, but little attention has been given to individual galaxy groups. We investigate the scaling relations of galaxy groups and clusters near the North Ecliptic Pole using X-ray and SZ observations. This region of the sky is where both the {\it{ROSAT}} and {\it{Planck}} satellites achieved their deepest observations, permitting the investigation of lower mass systems. Our sample consists of 62 X-ray detected groups and clusters, spanning a mass range of $10^{13.4}M_{\odot}<~M_{500}<10^{15}M_{\odot}$ and redshifts of $0.03\lesssim z \lesssim 0.82$. We extract the total SZ flux from unresolved {\it{Planck}} data and estimate the fraction of the SZ flux within $R_{500}$ assuming two different pressure profiles. The SZ scaling relations were derived using a Bayesian technique that accounts for censored data. We find a power law slope of $1.73^{+0.19}_{-0.18}$ for the $Y_{SZ}-M_{500}$ relation which is consistent with the self-similar prediction of 5/3. The slope of $0.89^{+0.09}_{-0.08}$ for the $Y_{SZ}-L_{X,500}$ relation is in agreement with other observational studies but not the self-similar prediction of 5/4, and the $Y_{SZ}-Y_{X}$ relation lies below the 1:1 relation when the slope is fixed to unity. The determined scaling relations are dependent on the selected pressure profile, so resolved data are needed to determine the effects of AGN feedback. In addition, we find a number of potential cluster candidates in the {\it{Planck}} Compton maps that were not identified in our X-ray sample.

\end{abstract}


\section{Introduction}

In the hierarchical growth model, low-mass galaxy groups agglomerate at high redshifts to form today's massive galaxy clusters with dark halo masses $\gtrsim 10^{14} M_{\odot}$ \citep[for a review on cluster formation see][]{Kravtsov12}. Although galaxy clusters are comprised of galaxy groups, their properties are very different, including: gas fractions, star formation efficiencies, and X-ray morphologies \citep{Mulchaey00,Dai10,Behroozi13,McCarthy17}. It is believed the total enclosed mass dictates these different qualities, meaning, there may be some characteristic mass where these systems become distinct \citep{Paul17}. However, groups and clusters exhibit a spectrum of masses so the boundary is somewhat vague.

Groups and clusters host large reservoirs of hot baryons surrounding their member galaxies. These gaseous halos, known as the intracluster medium (ICM), account for the majority of baryons in these systems \citep{Andreon10,Dai10,Kravtsov12}. (The ICM is referred to as the hot gaseous halo, regardless if the system is a group or a cluster.) The ICM is formed through the gravitational collapse of the intergalactic medium, during which gas is compressed and heated until it reaches hydrostatic equilibrium. Gravitational heating is a scale-free mechanism that only depends on the total enclosed mass. If gravitational collapse were the dominant processes in forming the ICM then it should scale self-similarly \citep{Kaiser86}. In reality, the ICM is constantly being affected by non-gravitational mechanisms such as accretion shocks and AGN feedback \citep{Bhattacharya08,Fabian12,Lau15}. 

The halos of galaxy groups are more susceptible to the effects of non-gravitational processes compared to their higher mass counterparts for a few reasons. First, groups reside in smaller potential wells, meaning non-gravitational processes may contribute a significant portion of the thermal energy compared to gravitational heating alone. Second, feedback from galactic winds and active galactic nuclei (AGNs) can create an entropy floor in groups by injecting high entropy gas and removing low entropy gas. This prevents ambient gas from accreting and, therefore, reduces the gas fraction \citep{Pratt09,McCarthy11,LeBrun17,Truong18}. Galaxy clusters, on the other hand, behave as ``closed-box'' systems and re-accrete any ejected material \citep{Farahi18}. Third, galaxy groups are thought to exist in a wide range of virialization states \citep{Mulchaey00,Balogh11} while most clusters have had enough time to relax. Lastly, groups are much more likely to experience mergers that cause morphological disruption. In the case of a dynamically young system, the assumptions of hydrostatic equilibrium and spherical symmetry become invalid, making it a challenge to characterize the ICM. Mergers can also produce shocks that heat the ICM.

Many efforts have been made in the X-rays to search for the differences between the halos of groups and clusters (see \citealt{Giodini13} for a review on cluster scaling relations). Simulations indicate there should be a change in the scaling relations when moving from groups to clusters. The predicted changes are those in the slopes and scatter at different mass scales, which are attributed to the differences in gas fractions and ranges of dynamical states \citep{Dave02, LeBrun17, Paul17, Farahi18}. Some observational studies claim to have found such evidence at low masses. For example, \citet{Eckmiller11} reported increases in scatter and \citet{Lovisari15} found changes in their slopes towards the group regime. On the other hand, massive clusters appear self-similar once their central cores ($R \lesssim 0.15~ R_{500}$) have been excised \citep{Pratt09,Mantz16}.

In addition to X-rays, recent studies have used the Sunyaev\textendash Zel'dovich (SZ) effect to characterize the halos of groups and clusters. The SZ effect is the energy boost given to cosmic microwave background photons via inverse Compton scattering and is quantified via the Compton parameter
\begin{equation}
y = \frac{\sigma_{T}}{m_{e}c^{2}} \int_{0}^{\infty} P_{e}(r) dl
\end{equation}
where $\sigma_{T}$ is the Thompson cross section, $m_{e}$ is the mass of an electron, $c$ is the speed of light, $P_{e}(r)$ is the pressure  of electrons as a function of radius where $P_{e}(r) = k_{B}n_{e}(r)T_{e}(r)$, $k_{B}$ is the Boltzmann constant, and $dl$ is integrated along a line-of-sight. Its linear dependence on density provides a mass-weighted measure, making it sensitive to the outskirts of the ICM. X-ray emission depends on the square of the electron density, yielding an emission-weighted measure, which is dominated by clumps and the central cores.

Recent SZ studies have confirmed that clusters are consistent with the self-similar prediction \citep{Bonamente08, Andersson11, Marrone12, PlanckClusterCounts}. Current sensitivities of the SZ signal, however, have permitted studies of only the most massive systems ($\gtrsim 10^{14} M_{\odot}$). This paper is among the first studies to investigate the SZ scaling relations using lower mass systems. We use a sample of groups and clusters detected near the north ecliptic pole (NEP) where the {\it{ROSAT}} and {\it{Planck}} satellites obtained their deepest observations. The rest of the paper is structured in the following way: \autoref{sec:sample} describes the NEP sample, \autoref{sec:methods} explains the methods used to extract the SZ signal, \autoref{sec:results} presents the results, \autoref{sec:analysis} provides an analysis of our results and a discussion of future work, and \autoref{sec:conclusion} gives a brief summary and highlights the main conclusions of the study. 

Throughout this paper we use the following cosmological parameters: $\mathrm{H_{0}~=~70~km~s^{-1}~Mpc^{-1}}$, $\mathrm{\Omega_{M}~=~0.3}$, $\mathrm{\Omega_{\Lambda}~=~0.7}$, and $\mathrm{E(z) = \sqrt{\Omega_{\Lambda} + \Omega_{M}(1 + z)^{3}}}$. We discuss regions where the enclosed mass density within a sphere is some factor, $\Delta$, above the critical density, $\rho_{c}$, such that $M_{\Delta} = \frac{4 \pi}{3} \Delta \rho_{c}R_{\Delta}^{3}$.
\section{NEP Sample\label{sec:sample}}

The NEP has been a very popular region of the sky to perform deep, contiguous surveys. In particular, the scanning pattern of the {\it{ROSAT}} satellite yielded its longest exposures near the ecliptic poles. The region $\mathrm{17^{h}15^{m}<\alpha_{2000}<18^{h}45^{m}}$ and $\mathrm{62^{\circ}<\delta_{2000}<71^{\circ}}$ was investigated by \citet{Henry06} and these authors provided a flux-limited sample of sources above $\sim$ $\mathrm{2~x~10^{-14}~erg~cm^{-2}~s^{-1}}$ in the 0.5\textendash 2.0 keV band. Of the 442 sources in their catalog, 63 galaxy clusters were identified, however, two of the clusters appear to be part of a single system (RX J1724.2+6956 and RX J1724.1+7000) and has been noted by these authors. (Their redshifts and temperatures are identical, and they are only separated by $\sim 4^{\prime}$ or [0.18 Mpc]. In this study, the properties of the brighter detection were used.) For galaxy groups and clusters, they provided useful quantities including: unabsorbed flux measurements in the 0.5\textendash 2.0 keV band, the X-ray luminosities within $R_{200}$ also measured in the 0.5\textendash 2.0 keV band, X-ray temperatures estimated from the low-redshift luminosity-temperature relation \citep{White97}, and redshifts from their optical identification program \citep{Gioia03}. The sources in this catalog stem from high-quality X-ray data. \citet{Henry06} required the detect count rate SNR$>$4 for each source such that they contained a sufficient collection of photons; the median number of counts was 91 and the minimum was 28.

The {\it{Planck}} satellite also rendered its deepest observations near the ecliptic poles \citep{PlanckMission, PlanckOverview, PlanckYMAP}. The four year mission scanned the sky using seven frequency channels, spanning 30-857 GHz. These data were then used to construct Compton maps ($y$-maps) by \citet{PlanckYMAP}. The various frequency maps were combined using two linear combination methods: the Needlet Independent Linear Combination (NILC) and the Modified Internal Linear Combination Algorithm (MILCA). These algorithms aimed to minimize the variance of the reconstructed maps while preserving the SZ signal.

The NILC maps used combined data both from the high and low frequency instruments (HFI and LFI) to capture contamination at various scales, but the MILCA maps only uses data from the HFI. In both cases, the HFI maps were smoothed to a common resolution of $10^{\prime}$. We believe the maps produced by the NILC do a better job at removing contaminating radio sources (described below) and exhibit lower residuals compared to the MILCA maps. We extracted the SZ flux from both maps and found that their values agreed within the statistical uncertainties. In this study, we used the NILC maps to determine the scaling relations.
\section{Extraction of the SZ Signal\label{sec:methods}}

\subsection{Sources of Contamination \label{sec:contamination}}
The $y$-maps suffer from various sources of galactic and extragalactic contamination. Galactic thermal dust emission is the main source of contamination at large angular scales while the Cosmic Infrared Background (CIB), radio sources, and infrared sources dominate at small angular scales \citep{PlanckYMAP}. Unresolved radio sources appear as strong negative peaks while infrared sources produce weak positive signals. Considering the flux from a radio source follows a decreasing power law with frequency, it produces an excess of power relative to the CMB at low frequencies whereas the SZ signal from a group or cluster causes a decrement \citep{Rubino03}. The SZ signal from a cluster or group is kept positive by subtracting the decrease in flux relative to the CMB at low frequencies \citep{PlanckYMAP}. Since strong radio sources produce an increase in flux at low frequencies, subtracting this from the CMB yields a negative value.

Many of these point sources were identified and cataloged by \citet{PlanckPS}. These can be accounted for in the $y$-maps by applying the point source mask provided by \citet{PlanckYMAP}. The point source mask removed most the contamination, but some of the polluting signal still leaked into the background. In addition, it was confirmed that many strong radio sources were not removed by the mask after searching the 20-cm catalog \citep{20cm}. In order to better account for contamination, we decided to identify and remove radio sources manually. 

The manual removal of radio sources was an effective method for the most conspicuous negative regions, but a population of more modest radio sources still remained unaccounted for. To understand the impact of these weaker sources, we searched the 20-cm radio catalog in a $12^{\circ}~\mathrm{x}~12^{\circ}$ field centered around the NEP (we call this the NEP grid herein). The catalog indicated that 50 sources were not identified by our manual inspection. We then tested to see if these objects produced a non-trivial amount of contamination. Their signals were extracted by taking the average value inside a circle centered on the source with a $20^{\prime}$ radius. The signal from each source was compared to its immediate background which was defined as an annulus with a radius $60^{\prime}$ after excluding the inner circle.

\begin{figure}[htbp]
   \includegraphics[width=0.45\textwidth]{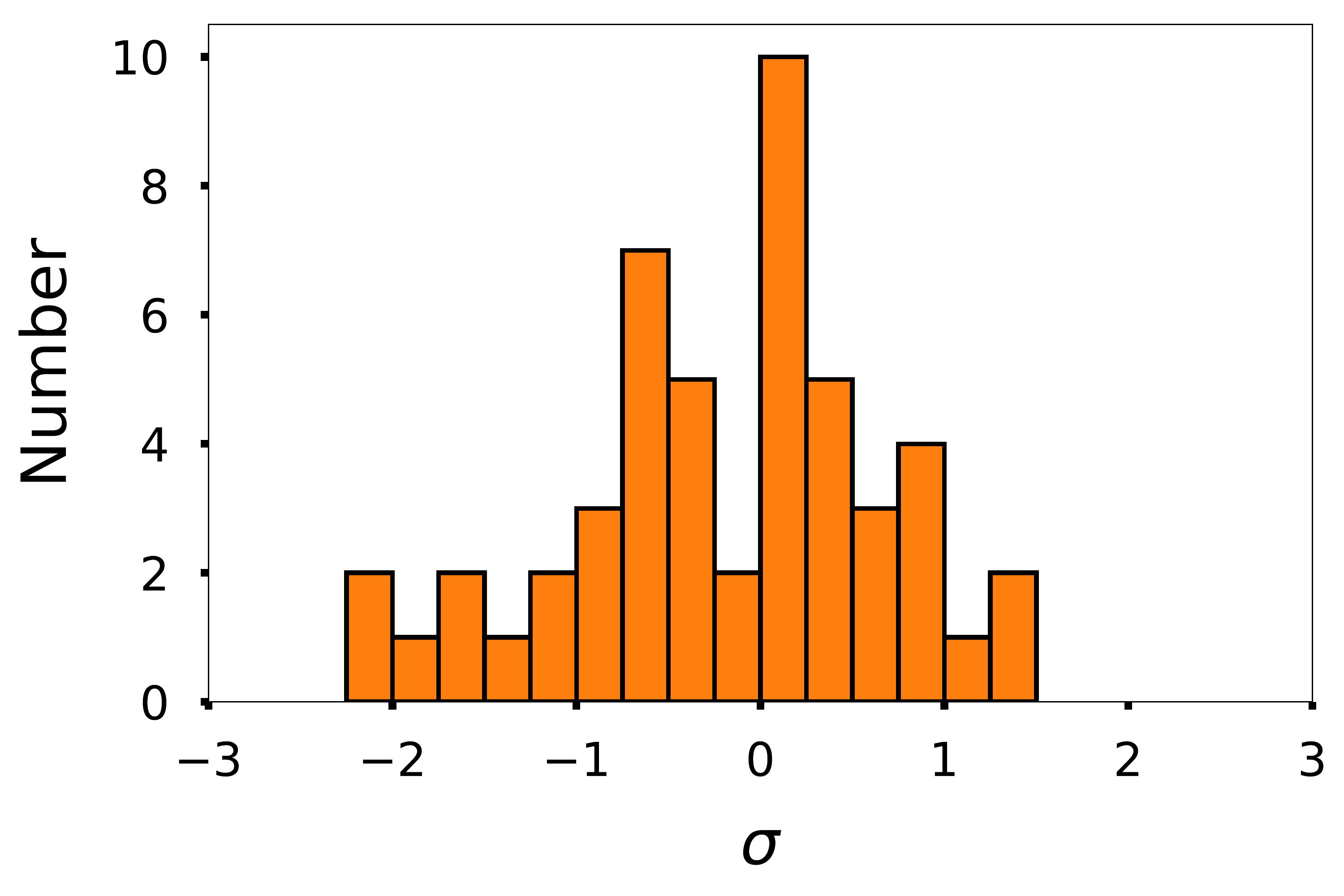}
    \caption{Histogram displaying the amount of contamination stemming from radio sources that were not identified in our manual removal program (see text). The $p$-values from the two-sided Student's t-test and the Wilcoxon signed-rank test were 0.13 and 0.10 respectively. These $p$-values are the probabilities of measuring a mean of zero purely by chance. A large $p$-value indicates one should not conclude that the mean of the distribution is something other than zero. A small $p$-value would these radio sources are stronger than their backgrounds and play a non-trivial role. The two tests yield $p > 0.05$, which is a canonical threshold used to determine a statistically significant result, suggesting the modest radio sources are not significant sources of contamination.}%
    \label{fig:radio_sigma_hist}%
\end{figure}

The differences between the radio sources and their backgrounds in units of standard error (see \autoref{sec:residuals}) are shown in \autoref{fig:radio_sigma_hist}. Radio sources should appear negative relative to their backgrounds in the $y$-maps i.e., on the left side of \autoref{fig:radio_sigma_hist}. Half of the sources yielded a positive signal relative to their backgrounds, and the mean of the distribution was $-0.18 \sigma$. We did not find strong enough statistical evidence, according to the two-sided t-test and the Wilcoxon signed-rank test, to reject the null hypothesis that the mean was consistent with zero. It was concluded that these modest radio sources did not have a significant effect in our ability to extract the SZ flux. The same result was obtained after combining the point source mask with our manual exclusion mask. Overall, there was no substantial improvement when implementing the point source mask, so it has not been included in this study. 

\subsection{Uncertainties \label{sec:residuals}}
There were two sources of error considered in this study: instrumental noise and the stochastic background. The stochastic background was quantified by the sample standard deviation in the Compton parameter, $\sigma_{y}$, as function of area. We created 360 sets of random annuli across the NEP grid where each annulus consisted of 18 bins separated by $10^{\prime}$. The five outermost bins determined the average local background which were then subtracted from the inner bins. The inner bins were used to quantify the background fluctuations by taking the average values from bins of comparable area to estimate $\sigma$ as a function of area. For example, the uncertainty in a $10^{\prime}$ circle was estimated by using the average values of the innermost bins across all sets of annuli. This was repeated for all annuli, giving $\sigma_{y}$ as a function of area. This can be expressed as
\begin{equation}
    \sigma_{y}(A) = \sqrt{\frac{\sum_{i=1}^{N} (p_{i}(A)-\bar{p}(A))^{2}}{N-1}}
\end{equation}
where $A$ is the area of common annuli, $p_{i}$ is the average pixel value in a single annulus, $\bar{p}$ is the average of all $p_{i}$, and $N$ is the number of alike annuli. 
\begin{figure}[htbp]
   \includegraphics[width=0.45\textwidth]{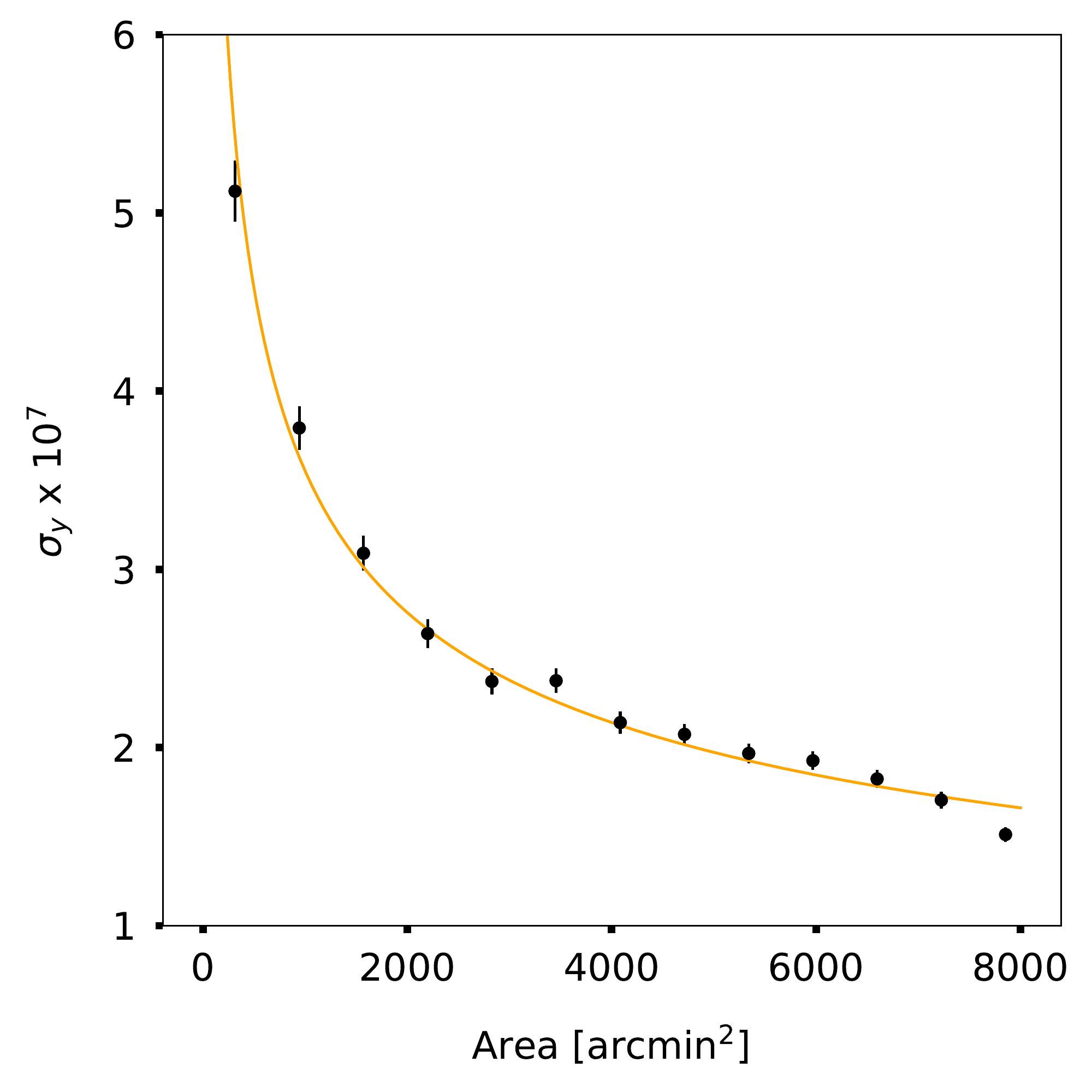}
    \caption{Best fit power law used to estimate the standard deviation in the Compton parameter as a function of area.}%
    \label{fig:sigma_area}%
\end{figure}

We point out that not all annuli were used in the analysis because many regions were obstructed by contaminating sources. There were many cases where most or all of the pixels within an annulus were unavailable due to nearby contaminants. In order to obtain robust estimates of the uncertainty, bins were required to retain $\geq 50\%$ of their pixels. We modeled $\sigma_{y}$ with a power law. We expected $\sigma_{y}$ to scale as the inverse square root of the area, however, our results yielded a shallower slope of -0.372 which is shown in \autoref{fig:sigma_area}. This flattened profile suggests there are large-scale secular variations in the $y$-maps.

Instrumental noise was characterized using the half difference maps provided by \citet{PlanckYMAP}. In these maps, the astrophysical emission has been eliminated, leaving only the instrumental component. The mean instrumental error was determined following a similar procedure for the stochastic background. The stochastic component was found to be $\sim 30\times$ larger than the instrumental component.


\subsection{Unresolved Sources}
For a typical cluster, most of the pressure of the ICM is found within \Rvir \citep{Arnaud10}. We estimated \Rvir for all of our systems using the $M_{500}-T_{X}$ relation for low mass clusters of galaxies from \citet{Kettula15}. On the $y$-maps, the projected radius is given by $\theta_{500} = \frac{R_{500}}{D_{A}}$ where $D_{A}$ is the angular distance. An object will be mostly unresolved if $\theta_{500}$ is less than the resolution of the $y$-maps ($10^{\prime}$). In our sample, only 13\% (8/62) of our objects have $\theta_{500} \geq 10^{'}$ while all of them have $\theta_{500}<21^{\prime}$. 

Since the SZ signals could not be resolved, we needed to understand the behavior of a point source and characterize the point spread function (PSF) of the $y$-maps. A sample of 27 bright point sources in the $y$-maps were used to determine the profile of the PSF. These were selected using the second Planck catalog of SZ sources \citep{PSZ2} in conjunction with the meta-catalog of X-ray detected clusters of galaxies \citep{Piffaretti11}. The clusters were selected using the following criteria: (1) $\theta_{500}<5^{\prime}$ to ensure they were unresolved (2) they were detected with high SNRs ($>10$) as quoted by \citet{PSZ2} (2) the centroids of the X-ray emission and SZ signal agreed within $5^{\prime}$ (4) the clusters did not appear to be involved in any obvious mergers based on the SZ images.

The PSF was modeled using a power law with a projected core radius of $\theta_{c}=10^{\prime}$. The profile takes the form 
\begin{equation}
\label{eq:psf}
{y_{psf}(\theta) = S_{0}\Big[1+\Big(\frac{\theta}{\theta_c}\Big)^2\Big]^{\frac{-3\beta}{2} + \frac{1}{2}} + S_{1}}
\end{equation}
where $S_{0}$ is the normalization, $S_1$ corrects for the local background, and $\beta$ characterizes the slope of the decay. $\beta$ was estimated using the profiles given the sample of known point sources. The point sources were sampled using 18 annuli in $10^{\prime}$ bins. Most of the signal was captured within the first 2\textendash 3 bins while the remaining bins determined the local background offset. We then determined the values of $\beta$ that minimized $\chi^{2}$ for each object. Our results yielded a mean slope of $2.39 \pm 0.15$ and a median of 2.29, which are consistent with each other. Using the median $\beta$ to characterize the PSF, we calculated the FWHM. Our PSF model yielded a FWHM of $10.3^{\prime}$, which is similar to the FWHM of $10^{\prime}$ used to smooth $y$-maps.

A $\beta$-model was arbitrarily chosen to be the functional form of the PSF. Another reasonable choice would be a Gaussian with zero mean. Following the same procedure as described above, one can estimate the width of the Gaussian for a point source. We found that a $\beta$-model systematically returns larger values compared to a Gaussian, albeit by a trivial 3.5 $\pm$ 0.1\%.

Some of sources were semi-resolved (i.e., $\theta_{500}>10^{'}$), so we modelled our data by convolving the PSF with the universal pressure profile (UPP) from \citet{Arnaud10} giving the functional form of the measured Compton parameter 
\begin{equation}\label{eq:convolve}
\tilde{y}(\theta) = y_{psf}(\theta)\circledast y_{upp}(\theta) 
\end{equation}
Assuming spherical symmetry, the total SZ flux, $Y_{SZ}$, in arcmin$^2$ is 
\begin{equation}
\label{eq: sz_flux}
Y_{SZ} = \int_{0}^{\infty} 2\pi \tilde{y}(\theta) \theta  d\theta
\end{equation}

The total flux can then be converted into the portion within a spherical region of radius \Rvir, but the unresolved nature of our objects required making assumptions about the pressure profiles. We considered two different pressure profiles to compute the fraction of the total SZ flux within \Rvir. The first method was to apply the UPP to all objects, which yields a constant fraction of $\approx 0.535$, which was found by integrating the UPP to \Rvir and to infinity. The second method was to apply the mass- and redshift-dependent pressure profiles provided by \citet{Battaglia10} (BPP herein), which takes into account the effects of AGN feedback. We denote the SZ flux within \Rvir as $Y_{sph,500}$ and the SZ luminosity as $Y_{sph,500}D_{A}^{2}$.


\subsection{Detections vs. Nondetections}
\begin{figure*}[htbp]
    \centering
    \includegraphics[width=0.9\textwidth]{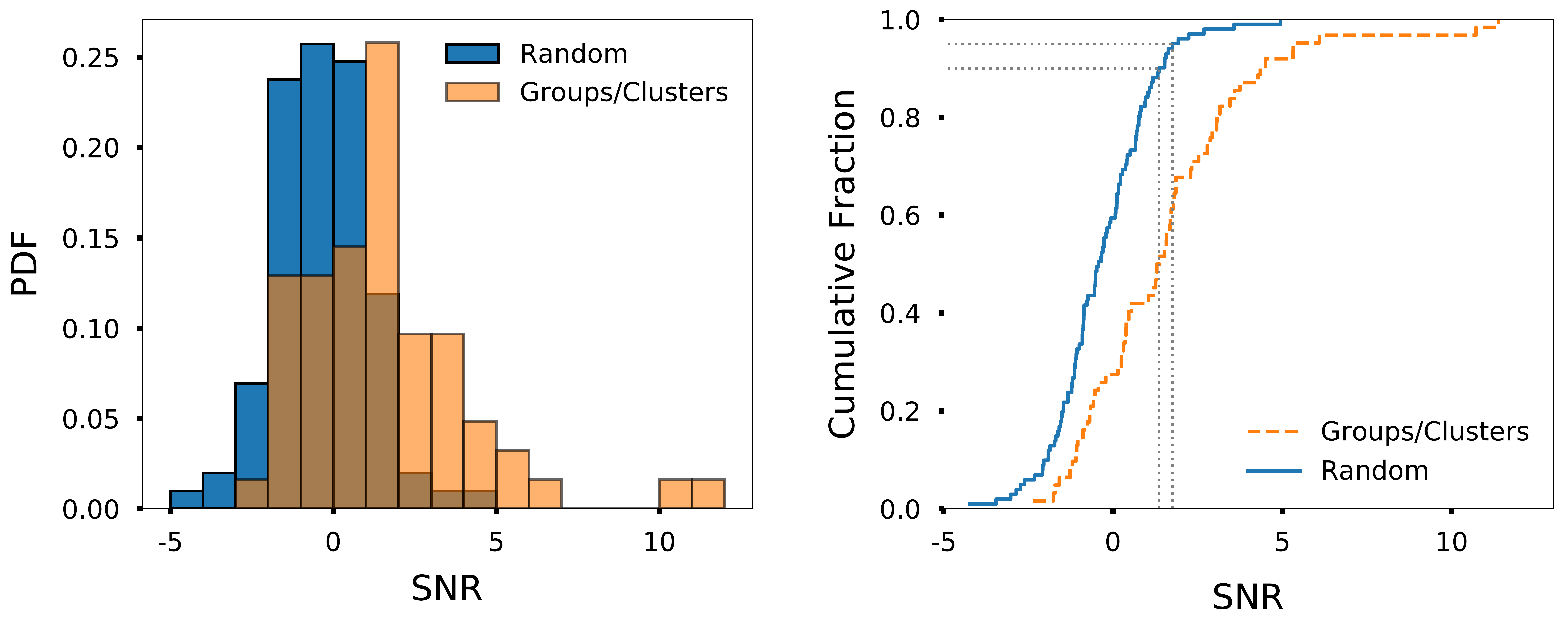} %
    \caption{SNR distributions for 62 groups/clusters and 101 random regions near the NEP. Included are normalized histograms (probability distribution function [PDF]) and cumulative fractions. The vertical and horizontal dashed lines in the right panels indicate where the cumulative fractions of SNRs for the random regions reach 90\% and 95\%.}%
    \label{fig:rand_henry}%
\end{figure*}
The sensitivity limit of the $y$-maps precluded us from extracting an acceptable SZ flux for every object, resulting in nondetections. In order to separate detections from nondetections, we investigated the set of random annuli in the NEP grid after removing sources of contamination. The annular bins were required to retain $\geq 50\%$ of their area in order to work with a robust sample. This was especially important for the innermost annuli ($\leq 30^{\prime}$) as they captured most of the flux from a point source. These data were fit with the PSF model that we used to calculate signal-to-noise ratios (SNRs) for each random region. We calculated the SNR as 
\begin{equation}
\label{eq:snr}
SNR = S_{0}/\sigma_{S_0}
\end{equation}
where $S_{0}$ is the normalization from \autoref{eq:psf} and $\sigma_{S_{0}}$ is the uncertainty on its estimated value. The same procedure was done for all of the known groups and clusters but with no restrictions placed upon the number of available pixels. Histograms and cumulative fractions of the SNR distributions are provided in \autoref{fig:rand_henry}. 

\autoref{fig:rand_henry} shows the SNR distributions did not stem from the same parent population and were statistically distinct according to the Anderson-Darling test ($p$-value $< 10^{-5}$). It also indicates where the background cumulative fractions reached 90\% and 95\%. The 90\% and 95\% limits occurred at SNR values of 1.35 and 1.75 respectively. These SNR values were considered as arbitrary thresholds where one could confidently detect a real SZ signal instead of a random fluctuation. We adopted the 90\% threshold (SNR $=$ 1.35) for this study, yielding 32 detections and 30 nondetections. Even though the choice of the SNR$>1.35$ threshold was arbitrary, we found that changing the limit to SNR$>1.75$ had negligible effects on the derived scaling relations which are presented in \autoref{sec:scalings}.
\section{Scaling Relations}\label{sec:results}

\subsection{X-ray Properties}

The X-ray data used in this study came from the NEP catalog constructed by \citet{Henry06}.  These authors measured photons count rates which were then converted into unabsorbed fluxes and luminosities within $R_{200}$. In order to stay consistent with our choice of \Rvir, we transformed the luminosity values to that inside \Rvir, denoted as $L_{X,500}$. This was done by taking the fraction of the emission measure inside \Rvir compared to that inside $R_{200}$ using an isothermal $\beta$ model for the electron density distribution 
\begin{equation}
n_{e}(r) = n_{e,0}\Big[1+\Big(\frac{r}{r_{c}}\Big)^{2}\Big]^{\frac{-3\beta}{2}}
\end{equation}
assuming $\beta=2/3$, $r_{c} = 0.15 ~R_{500}$, and $R_{500} = 0.7~ R_{200}$. This is simply a ratio of integrating $n_{e}^{2}$ over the two volumes, yielding a fraction $\approx 0.938$ for all objects.

We also computed the X-ray analog of the SZ luminosity within \Rvir, denoted as $Y_{X,500}$, which is defined as the product of the gas mass and the temperature \citep{Kravtsov06}. We estimated the gas mass using the APEC model simulation of a ROSAT PSPC spectrum. Inputting temperatures, abundances (assumed to be 0.3 Solar), redshifts, and exposure times as fixed parameters, the model predicted an energy flux, $F_{APEC}$, which was compared to the observed fluxes of our objects, $F_{obs}$. The ratio of these two fluxes were set equal to the normalization of the APEC model such that
\begin{equation}
\label{eq:APEC}
\frac{F_{obs}}{F_{APEC}} = \frac{10^{-14}}{4 \pi [D_{A}(1+z)]^{2}} \int n_{e}n_{H}dV
\end{equation}
where $D_{A}$ is measured in cm and $n_{e,H}$ is in $\mathrm{cm^{-3}}$.
We again assumed an isothermal $\beta$ model for the density distribution. We adopted the values used by \citet{Andersson11} for the mass density $\rho_{gas} = m_{p}n_{p}\mathrm{A}$ where the average nuclear mass was $\mathrm{A} = 1.397$ and $n_{e} = 1.199 n_{H}$. In order to compare $Y_{X,500}$ with $Y_{sph,500}$ there was a factor $\sigma_{T}/m_{e}c^{2}\mu_{e}m_{p} = 1.395 ~\times ~10^{-19} \mathrm{\frac{Mpc^{2}}{M_{\odot} keV}}$ to convert $Y_{X,500}$ into units of $\mathrm{Mpc^{2}}$. Values of $Y_{X,500}$ are provided in \autoref{tab:henry_info}. 


\subsection{Regression Methods \label{sec:regression_methods}}
Scaling relations are best modeled using linear regression in log-log space since the distributions are skewed in linear space. For example, the mass distribution in linear space is heavily clustered toward lower masses with only a few objects at the high-mass end. Fitting for the distribution in linear space would give the high-mass data significant leverage on the fit compared to the rest. In turn, we fit our data using the functional form $\mathrm{log_{10}}(Y/Y_{0}) = \alpha + \beta \mathrm{log_{10}}(X/X_{0})$ where $Y_{0}$ and $X_{0}$ are pivot points. The pivot points for $M_{500}$, $Y_{sph,500}D_{A}^{2}$, $Y_{X,500}$, and $L_{X}$, were $10^{14}~M_{\odot}$, $10^{-5}~\mathrm{Mpc^{2}}$, $10^{-5}~\mathrm{Mpc^{2}}$, and $10^{43} \mathrm{erg~s^{-1}}$ respectively.

The uncertainty in some variable {\it{X}} was approximated in log space as $\sigma_{log,X_{i}} = log_{10}(e) \sigma_{X_{i}}/X_{i}$ where $\sigma_{X_{i}}$ is the measured uncertainty on $X_{i}$ in linear space; this is a common practice in the literature \citep[e.g.,][]{Bonamente08, Eckmiller11,Lovisari15}. The error bars were slightly asymmetric in M$_{500}$ given the uncertainty and scatter in the M$_{500}-T_{X}$ relation from \citet{Kettula15}. These errors were approximated as symmetric using the mean error. The error bars on L$_{X,500}$ were scaled from the photon count rates and their uncertainties provided by \citet{Henry06} such that $\mathrm{\frac{L_{X,500}}{\delta L_{X,500}} = \frac{\Gamma}{\delta \Gamma}}$ where $\Gamma$ represents the photon count rate and $\delta$ represents the uncertainty of a variable. The uncertainties on $Y_{X,500}$ were dominated by the $64\%$ intrinsic scatter adopted from the results of the  $Y_{X}-L_{X}$ relation determined by \citet{Lovisari15}.
\startlongtable
\begin{deluxetable*}{llllrccrrr}
\tabletypesize{\footnotesize}
\tablecaption{Measured X-ray and SZ Values \label{tab:henry_info}}
\tablehead{\colhead{ID} & \colhead{$\alpha$ (J2000)} & \colhead{$\delta$ (J2000)} & \colhead{z} & \colhead{L$_{\mathrm{X,500}}$} & \colhead{T} & \colhead{M$_{\mathrm{500}}$} & \colhead{Y$\mathrm{_{X,500}}$} & \colhead{SNR} &  \colhead{Y$\mathrm{_{SZ}}$} \\\colhead{} & \colhead{} & \colhead{} & \colhead{} & \colhead{[$10^{42}~ \mathrm{erg ~s^{-1}}$]}& \colhead{[keV]} & \colhead{[$10^{14}$ M$_{\odot}$]} & \colhead{[$10^{-6}$ Mpc$^{2}$]}& \colhead{} & \colhead{[$10^{-4}$ arcmin$^{2}$]}}
\colnumbers
\startdata
RX J1716.6+6410 & 17:16:39.7 & 64:10:35 & 0.2507 & 43.1 $\pm$ 6.6 & 3.7 & 3.13 & 10.02 $\pm$ 2.43 & 4.63 & 11.21 $\pm$ 2.42 \\
RX J1721.4+6733 & 17:21:24.6 & 67:33:14 & 0.0861 & 1.87 $\pm$ 0.38 & 1.3 & 0.70 & 0.36 $\pm$ 0.09 & -0.59 & -0.67 $\pm$ 1.15 \\
RX J1724.2+6956 & 17:24:16.0 & 69:56:44 & 0.0386 & 0.861 $\pm$ 0.2 & 0.9 & 0.41 & 0.11 $\pm$ 0.03 & 0.39 & 0.47 $\pm$ 1.20 \\
RX J1724.7+6716 & 17:24:47.3 & 67:16:09 & 0.2540 & 15 $\pm$ 3.4 & 2.6 & 1.83 & 3.14 $\pm$ 0.81 & 1.82 & 2.39 $\pm$ 1.31 \\
RX J1727.4+7035 & 17:27:25.8 & 70:35:37 & 0.3059 & 110 $\pm$ 14 & 5.5 & 5.54 & 31.39 $\pm$ 7.52 & 3.17 & 7.44 $\pm$ 2.34 \\
RX J1728.6+7041 & 17:28:39.5 & 70:41:05 & 0.5509 & 227 $\pm$ 41 & 7.3 & 7.39 & 60.72 $\pm$ 15.03 & 3.24 & 9.68 $\pm$ 2.99 \\
RX J1735.0+6405 & 17:35:04.9 & 64:05:57 & 0.1411 & 44 $\pm$ 3.6 & 3.7 & 3.32 & 11.08 $\pm$ 2.59 & 1.19 & 2.99 $\pm$ 2.52 \\
RX J1736.3+6802 & 17:36:23.4 & 68:02:06 & 0.0258 & 2.81 $\pm$ 0.21 & 1.4 & 0.80 & 0.50 $\pm$ 0.12 & 3.60 & 12.10 $\pm$ 3.36 \\
RX J1742.7+6735 & 17:42:46.8 & 67:35:53 & 0.0420 & 2.81 $\pm$ 0.28 & 1.3 & 0.71 & 0.39 $\pm$ 0.09 & 3.57 & 10.35 $\pm$ 2.90 \\
RX J1743.3+6440 & 17:43:23.3 & 64:40:18 & 0.1790 & 27.1 $\pm$ 3.1 & 3.2 & 2.61 & 6.48 $\pm$ 1.54 & 3.02 & 5.75 $\pm$ 1.91 \\
RX J1743.4+6341 & 17:43:28.1 & 63:41:39 & 0.3270 & 125 $\pm$ 11 & 5.7 & 5.78 & 34.98 $\pm$ 8.22 & 4.30 & 9.21 $\pm$ 2.14 \\
RX J1745.2+6556 & 17:45:16.2 & 65:56:17 & 0.6080 & 74.9 $\pm$ 16 & 4.5 & 3.43 & 13.79 $\pm$ 3.51 & -2.36 & -3.60 $\pm$ 1.53 \\
RX J1746.7+6639 & 17:46:45.0 & 66:39:20 & 0.3864 & 33.7 $\pm$ 7 & 3.4 & 2.55 & 6.82 $\pm$ 1.72 & 0.23 & 0.33 $\pm$ 1.40 \\
RX J1747.5+6343 & 17:47:33.6 & 63:43:55 & 0.3280 & 40.3 $\pm$ 7.9 & 3.6 & 2.88 & 8.67 $\pm$ 2.17 & 3.72 & 7.75 $\pm$ 2.08 \\
RX J1748.6+7020 & 17:48:41.6 & 70:20:31 & 0.3450 & 40.3 $\pm$ 7.7 & 3.6 & 2.85 & 8.60 $\pm$ 2.14 & -1.24 & -5.11 $\pm$ 4.11 \\
RX J1749.0+7014 & 17:49:03.5 & 70:14:42 & 0.5790 & 215 $\pm$ 29 & 7.1 & 6.97 & 54.88 $\pm$ 13.19 & -0.82 & -6.15 $\pm$ 7.50 \\
RX J1749.8+6823 & 17:49:49.8 & 68:23:15 & 0.0508 & 0.955 $\pm$ 0.16 & 1.0 & 0.47 & 0.14 $\pm$ 0.03 & 1.93 & 4.24 $\pm$ 2.20 \\
RX J1751.2+6533 & 17:51:15.5 & 65:33:33 & 0.0424 & 0.674 $\pm$ 0.11 & 0.9 & 0.41 & 0.09 $\pm$ 0.02 & 1.45 & 3.27 $\pm$ 2.26 \\
RX J1751.5+7013 & 17:51:30.7 & 70:13:32 & 0.4925 & 83.3 $\pm$ 16 & 4.9 & 4.18 & 18.82 $\pm$ 4.70 & -1.13 & -3.76 $\pm$ 3.33 \\
RX J1751.5+6719 & 17:51:30.9 & 67:19:20 & 0.0933 & 3.74 $\pm$ 0.4 & 1.6 & 0.95 & 0.73 $\pm$ 0.17 & 5.59 & 7.95 $\pm$ 1.42 \\
RX J1752.2+6522 & 17:52:12.0 & 65:22:22 & 0.3923 & 26.2 $\pm$ 5.2 & 3.0 & 2.10 & 4.78 $\pm$ 1.20 & 0.55 & 0.71 $\pm$ 1.30 \\
RX J1754.5+6904 & 17:54:35.0 & 69:04:58 & 0.5113 & 53.4 $\pm$ 11 & 4.0 & 3.03 & 10.28 $\pm$ 2.61 & 2.33 & 7.41 $\pm$ 3.18 \\
RX J1754.6+6803 & 17:54:41.9 & 68:03:33 & 0.0770 & 34.6 $\pm$ 0.98 & 3.4 & 3.01 & 8.86 $\pm$ 2.04 & 15.44 & 40.92 $\pm$ 2.65 \\
RX J1754.7+6623 & 17:54:45.7 & 66:23:53 & 0.0879 & 2.81 $\pm$ 0.19 & 1.5 & 0.86 & 0.60 $\pm$ 0.14 & -1.68 & -4.01 $\pm$ 2.38 \\
RX J1754.0+6452 & 17:54:05.3 & 64:52:01 & 0.2460 & 11.2 $\pm$ 2.3 & 2.4 & 1.62 & 2.43 $\pm$ 0.61 & 1.71 & 4.03 $\pm$ 2.35 \\
RX J1755.3+6504 & 17:55:19.9 & 65:04:55 & 0.0846 & 1.87 $\pm$ 0.27 & 1.2 & 0.62 & 0.28 $\pm$ 0.07 & 0.38 & 0.92 $\pm$ 2.41 \\
RX J1755.7+6752 & 17:55:45.5 & 67:52:42 & 0.0833 & 13.1 $\pm$ 0.7 & 2.5 & 1.88 & 3.17 $\pm$ 0.73 & 6.33 & 17.92 $\pm$ 2.83 \\
RX J1755.8+6236 & 17:55:48.3 & 62:36:41 & 0.0270 & 3.74 $\pm$ 0.24 & 1.5 & 0.89 & 0.64 $\pm$ 0.15 & 1.94 & 8.71 $\pm$ 4.48 \\
RX J1755.9+6314 & 17:55:56.5 & 63:14:03 & 0.3850 & 63.7 $\pm$ 9.5 & 4.2 & 3.52 & 13.67 $\pm$ 3.31 & 1.54 & 3.62 $\pm$ 2.36 \\
RX J1756.5+6513 & 17:56:31.0 & 65:13:01 & 0.0284 & 0.571 $\pm$ 0.068 & 0.8 & 0.34 & 0.07 $\pm$ 0.02 & -1.02 & -2.23 $\pm$ 2.18 \\
RX J1757.3+6631 & 17:57:19.8 & 66:31:39 & 0.6909 & 56.2 $\pm$ 7.2 & 4.1 & 2.83 & 9.38 $\pm$ 2.24 & -0.44 & -1.20 $\pm$ 2.72 \\
RX J1758.9+6520 & 17:58:57.6 & 65:20:58 & 0.3652 & 15.9 $\pm$ 3.2 & 2.6 & 1.72 & 2.93 $\pm$ 0.73 & -0.59 & -1.14 $\pm$ 1.92 \\
RX J1759.2+6902 & 17:59:17.5 & 69:02:20 & 0.0994 & 2.81 $\pm$ 0.47 & 1.5 & 0.86 & 0.57 $\pm$ 0.14 & 3.71 & 8.80 $\pm$ 2.37 \\
RX J1800.4+6913 & 18:00:28.2 & 69:13:22 & 0.0821 & 14 $\pm$ 0.85 & 2.6 & 2.00 & 3.51 $\pm$ 0.82 & 0.40 & 1.06 $\pm$ 2.63 \\
RX J1811.3+6447 & 18:11:19.1 & 64:47:36 & 0.4510 & 71.1 $\pm$ 12 & 4.4 & 3.63 & 14.88 $\pm$ 3.63 & -1.06 & -3.10 $\pm$ 2.94 \\
RX J1812.1+6353 & 18:12:08.4 & 63:53:35 & 0.5408 & 132 $\pm$ 21 & 5.9 & 5.38 & 31.94 $\pm$ 7.79 & 0.25 & 0.57 $\pm$ 2.26 \\
RX J1813.1+6230 & 18:13:11.5 & 62:30:33 & 0.1829 & 12.2 $\pm$ 2.8 & 2.5 & 1.79 & 2.83 $\pm$ 0.73 & 1.42 & 2.16 $\pm$ 1.53 \\
RX J1814.2+6939 & 18:14:14.4 & 69:39:33 & 0.0874 & 13.1 $\pm$ 1.1 & 2.5 & 1.88 & 3.17 $\pm$ 0.74 & 2.44 & 3.99 $\pm$ 1.63 \\
RX J1816.5+6911 & 18:16:32.4 & 69:11:34 & 0.2097 & 13.1 $\pm$ 2.7 & 2.5 & 1.76 & 2.89 $\pm$ 0.73 & 1.22 & 2.18 $\pm$ 1.78 \\
RX J1817.7+6824 & 18:17:46.1 & 68:24:24 & 0.2820 & 137 $\pm$ 9 & 5.9 & 6.25 & 40.68 $\pm$ 9.46 & 1.46 & 3.17 $\pm$ 2.17 \\
RX J1817.1+7024 & 18:17:08.4 & 70:24:13 & 0.0859 & 3.74 $\pm$ 0.73 & 1.5 & 0.86 & 0.61 $\pm$ 0.15 & 0.26 & 0.55 $\pm$ 2.13 \\
RX J1819.0+6909 & 18:19:04.1 & 69:09:24 & 0.0880 & 3.74 $\pm$ 0.73 & 1.6 & 0.95 & 0.72 $\pm$ 0.18 & 1.32 & 3.45 $\pm$ 2.62 \\
RX J1819.8+6748 & 18:19:48.8 & 67:48:48 & 0.2153 & 14 $\pm$ 2.9 & 2.6 & 1.87 & 3.17 $\pm$ 0.80 & 0.98 & 1.10 $\pm$ 1.13 \\
RX J1802.9+6339 & 18:02:54.0 & 63:39:10 & 0.0907 & 1.22 $\pm$ 0.29 & 1.1 & 0.54 & 0.19 $\pm$ 0.05 & -0.73 & -1.75 $\pm$ 2.37 \\
RX J1820.2+6857 & 18:20:13.0 & 68:57:22 & 0.0890 & 25.3 $\pm$ 1.5 & 3.0 & 2.47 & 5.99 $\pm$ 1.39 & 1.64 & 4.03 $\pm$ 2.45 \\
RX J1821.6+6827 & 18:21:38.1 & 68:27:52 & 0.8156 & 198 $\pm$ 43 & 6.9 & 5.79 & 40.36 $\pm$ 10.29 & 1.57 & 3.48 $\pm$ 2.22 \\
RX J1822.6+6641 & 18:22:37.4 & 66:41:29 & 0.0888 & 2.81 $\pm$ 0.46 & 1.4 & 0.78 & 0.45 $\pm$ 0.11 & -1.91 & -3.56 $\pm$ 1.86 \\
RX J1829.0+6913 & 18:29:03.7 & 69:13:50 & 0.2057 & 88.9 $\pm$ 7.7 & 5.0 & 5.07 & 25.74 $\pm$ 6.03 & 1.76 & 3.87 $\pm$ 2.20 \\
RX J1832.2+6832 & 18:32:13.3 & 68:32:26 & 0.1981 & 18.7 $\pm$ 3.9 & 2.8 & 2.11 & 4.17 $\pm$ 1.05 & 1.74 & 3.12 $\pm$ 1.79 \\
RX J1832.5+6449 & 18:32:31.5 & 64:49:49 & 0.1610 & 73 $\pm$ 4.1 & 4.6 & 4.57 & 20.71 $\pm$ 4.81 & 12.45 & 20.15 $\pm$ 1.62 \\
RX J1832.5+6848 & 18:32:35.0 & 68:48:05 & 0.2050 & 241 $\pm$ 11 & 7.4 & 9.20 & 86.38 $\pm$ 20.00 & 2.80 & 4.14 $\pm$ 1.48 \\
RX J1833.7+6521 & 18:33:44.6 & 65:21:37 & 0.1621 & 19.7 $\pm$ 2.3 & 3.0 & 2.39 & 5.07 $\pm$ 1.20 & -1.77 & -2.50 $\pm$ 1.41 \\
RX J1834.1+7057 & 18:34:08.2 & 70:57:23 & 0.0803 & 9.36 $\pm$ 1 & 2.3 & 1.66 & 2.30 $\pm$ 0.55 & 3.85 & 12.77 $\pm$ 3.31 \\
RX J1836.5+6344 & 18:36:31.0 & 63:44:30 & 0.0846 & 34.6 $\pm$ 1.8 & 3.4 & 3.00 & 8.80 $\pm$ 2.04 & 7.07 & 11.70 $\pm$ 1.66 \\
RX J1838.2+6321 & 18:38:12.6 & 63:21:02 & 0.2167 & 43.1 $\pm$ 5.1 & 3.7 & 3.19 & 10.32 $\pm$ 2.45 & -1.30 & -1.94 $\pm$ 1.49 \\
RX J1839.2+7018 & 18:39:17.4 & 70:18:20 & 0.2297 & 18.7 $\pm$ 4.5 & 2.8 & 2.07 & 4.12 $\pm$ 1.07 & -1.04 & -2.87 $\pm$ 2.77 \\
RX J1804.2+6729 & 18:04:15.6 & 67:29:21 & 0.0617 & 0.431 $\pm$ 0.091 & 0.7 & 0.27 & 0.05 $\pm$ 0.01 & 0.48 & 0.75 $\pm$ 1.57 \\
RX J1806.4+7028 & 18:06:24.9 & 70:28:40 & 0.0971 & 7.49 $\pm$ 0.93 & 2.1 & 1.43 & 1.75 $\pm$ 0.42 & 0.29 & 0.60 $\pm$ 2.09 \\
RX J1806.8+6537 & 18:06:51.6 & 65:37:46 & 0.2626 & 49.6 $\pm$ 3.5 & 3.9 & 3.37 & 11.77 $\pm$ 2.74 & 4.33 & 11.22 $\pm$ 2.59 \\
RX J1806.1+6813 & 18:06:06.6 & 68:13:08 & 0.3030 & 43.1 $\pm$ 4.9 & 3.7 & 3.04 & 9.66 $\pm$ 2.29 & 3.10 & 2.37 $\pm$ 0.77 \\
RX J1807.5+6429 & 18:07:32.3 & 64:29:17 & 0.2391 & 11.2 $\pm$ 2.4 & 2.4 & 1.63 & 2.39 $\pm$ 0.61 & -0.24 & -0.58 $\pm$ 2.41 \\
RX J1808.7+6557 & 18:08:43.6 & 65:57:05 & 0.2460 & 9.36 $\pm$ 1.4 & 2.3 & 1.52 & 2.07 $\pm$ 0.50 & -0.88 & -2.17 $\pm$ 2.45 \\
\enddata
\tablecomments{The first four columns and column 6 are from \citet{Henry06}. Column 5 is the X-ray luminosity in 0.5-2.0 keV band within \Rvir (see text). Column 7 is the halo mass within \Rvir estimated from the $M_{500}-T_{X}$ relation from  \citet{Kettula15} given the temperature in column 6. Column 8 is the X-ray analog  of the SZ luminosity within \Rvir. Column 9 is the SZ signal-to-noise ratio calculated  using \autoref{eq:snr}. Column 10 is the total SZ flux computed from \autoref{eq: sz_flux}.}
\end{deluxetable*}
\vspace{2cm}


\begin{deluxetable*}{cccccccc}[t]
\tablewidth{0pt} 
\tablecaption{Scaling Relations using the \citet{Arnaud10} Pressure Profile
\label{tab:regression_arnaud}}
\tablehead{
\colhead{} & \colhead{} & \colhead{} & \colhead{} & \colhead{} & \multicolumn{3}{c}{Bias Corrected}
\\
\cline{6-8}
\colhead{Relation ($Y$ - $X$)} & \colhead{Method}  & \colhead{$\alpha$} & \colhead{$\beta$} & \colhead{$\sigma_{log,int}$}  & \colhead{$\alpha$} & \colhead{$\beta$} & \colhead{$\sigma_{log,int}$}}
\startdata 
\\
{$\mathrm{Y_{sph,500}D_{A}^{2}E(z)^{-2/3}}$ - $\mathrm{M_{500}}$} & {$\mathrm{Bayesian}$} & $-0.61^{+0.10}_{-0.10}$ & $1.37^{+0.20}_{-0.20}$ & $0.30^{+0.06}_{-0.05}~ (70\%)$ & $-0.65^{+0.11}_{-0.11}$ & $1.32^{+0.21}_{-0.21}$ & $0.35^{+0.07}_{-0.06}~ (81\%)$ \\ \\
 {} & {$\mathrm{Orthogonal}$}  & $\mathrm{-0.78^{+0.09}_{-0.09}}$ & $\mathrm{1.77^{+0.20}_{-0.20}}$ & $\mathrm{0.20^{+0.03}_{-0.03}~(46\%)}$ & $\mathrm{-0.86^{+0.09}_{-0.09}}$ & $\mathrm{1.85^{+0.21}_{-0.21}}$ & $\mathrm{0.32^{+0.03}_{-0.03}~ (73\%)}$ \\ \\
\hline \\
{$\mathrm{Y_{sph,500}D_{A}^{2}E(z)^{-2/3}}$ - $\mathrm{L_{X,500}E(z)^{-7/3}}$} & {$\mathrm{Bayesian}$} & $-0.30^{+0.07}_{-0.07}$ & $0.74^{+0.11}_{-0.11}$ & $0.34^{+0.06}_{-0.05}~ (77\%)$ & $-0.33^{+0.08}_{-0.08}$ & $0.73^{+0.12}_{-0.12}$ & $0.37^{+0.07}_{-0.05}~ (85\%)$ \\ \\
 {} & {$\mathrm{Orthogonal}$} & $\mathrm{-0.35^{+0.07}_{-0.07}}$ & $\mathrm{0.89^{+0.11}_{-0.11}}$ & $\mathrm{0.24^{+0.06}_{-0.06}~(56\%)}$ & $\mathrm{-0.40^{+0.07}_{-0.07}}$ & $\mathrm{0.91^{+0.12}_{-0.12}}$ & $\mathrm{0.35^{+0.01}_{-0.01}~ (81\%)}$ \\ \\
\hline \\
{$\mathrm{Y_{sph,500}D_{A}^{2}}$ - $\mathrm{Y_{X,500}}$} & {$\mathrm{Bayesian}$} & $\mathrm{0.06^{+0.07}_{-0.08}}$ & $\mathrm{1 ~ (fixed)}$ & $\mathrm{0.39^{+0.06}_{-0.05}~ (90\%)}$ & $\mathrm{-0.01^{+0.08}_{-0.08}}$ & $\mathrm{1 ~ (fixed)}$ & $\mathrm{0.44^{+0.07}_{-0.06}~ (102\%)}$ \\ \\
{} & {$\mathrm{Orthogonal}$} & $\mathrm{0.19^{+0.08}_{-0.08}}$ & $\mathrm{1 ~ (fixed)}$ & $\mathrm{0.28^{+0.03}_{-0.03}~ (65\%)}$ & $\mathrm{0.15^{+0.09}_{-0.09}}$ & $\mathrm{1 ~ (fixed)}$ & $\mathrm{0.33^{+0.02}_{-0.02}~ (77\%)}$ \\ \\
\enddata
\tablecomments{Estimated parameters for various scaling relations under the assumption that all systems follow the ``universal pressure profile'' from \citet{Arnaud10}.}
\end{deluxetable*}
\begin{deluxetable*}{cccccccc}[!t]
\tablewidth{0pt} 
\tablecaption{Scaling Relations using the \citet{Battaglia10} Pressure Profile
\label{tab:regression_battaglia}}
\tablehead{
\colhead{} & \colhead{} & \colhead{} & \colhead{} & \colhead{} & \multicolumn{3}{c}{Bias Corrected}
\\
\cline{6-8}
\colhead{Relation ($Y$ - $X$)} & \colhead{Method}  & \colhead{$\alpha$} & \colhead{$\beta$} & \colhead{$\sigma_{log,int}$}  & \colhead{$\alpha$} & \colhead{$\beta$} & \colhead{$\sigma_{log,int}$}}
\startdata 
\\
{$\mathrm{Y_{sph,500}D_{A}^{2}E(z)^{-2/3}}$ - $\mathrm{M_{500}}$} & {$\mathrm{Bayesian}$} & $-0.89^{+0.08}_{-0.09}$ & $1.75^{+0.17}_{-0.17}$ & $0.30^{+0.06}_{-0.05}~ (70\%)$ & $-0.95^{+0.09}_{-0.10}$ & $1.73^{+0.19}_{-0.18}$ & $0.33^{+0.06}_{-0.06}~ (77\%)$ \\ \\
 {} & {$\mathrm{Orthogonal}$}  & $\mathrm{-0.91^{+0.09}_{-0.09}}$ & $\mathrm{2.03^{+0.20}_{-0.20}}$ & $\mathrm{0.16^{+0.03}_{-0.03}~(37\%)}$ & $\mathrm{-0.98^{+0.09}_{-0.09}}$ & $\mathrm{2.09^{+0.20}_{-0.20}}$ & $\mathrm{0.27^{+0.02}_{-0.02}~ (62\%)}$ \\ \\
\hline \\
{$\mathrm{Y_{sph,500}D_{A}^{2}E(z)^{-2/3}}$ - $\mathrm{L_{X,500}E(z)^{-7/3}}$} & {$\mathrm{Bayesian}$} & $-0.56^{+0.06}_{-0.07}$ & $0.90^{+0.08}_{-0.08}$ & $0.31^{+0.05}_{-0.04}~ (72\%)$ & $-0.62^{+0.07}_{-0.07}$ & $0.89^{+0.09}_{-0.08}$ & $0.34^{+0.06}_{-0.05}~ (78\%)$ \\ \\
 {} & {$\mathrm{Orthogonal}$} & $\mathrm{-0.42^{+0.07}_{-0.07}}$ & $\mathrm{1.05^{+0.12}_{-0.12}}$ & $\mathrm{0.24^{+0.07}_{-0.07}~(56\%)}$ & $\mathrm{-0.48^{+0.07}_{-0.07}}$ & $\mathrm{1.07^{+0.12}_{-0.12}}$ & $\mathrm{0.34^{+0.02}_{-0.02}~ (79\%)}$ \\ \\
\hline \\
{$\mathrm{Y_{sph,500}D_{A}^{2}}$ - $\mathrm{Y_{X,500}}$} & {$\mathrm{Bayesian}$} & $\mathrm{0.04^{+0.06}_{-0.07}}$ & $\mathrm{1 ~ (fixed)}$ & $\mathrm{0.35^{+0.06}_{-0.05}~ (81\%)}$ & $\mathrm{-0.02^{+0.07}_{-0.08}}$ & $\mathrm{1 ~ (fixed)}$ & $\mathrm{0.39^{+0.06}_{-0.05}~ (90\%)}$ \\ \\ 
{} & {$\mathrm{Orthogonal}$} & $\mathrm{0.15^{+0.07}_{-0.07}}$ & $\mathrm{1 ~ (fixed)}$ & $\mathrm{0.17^{+0.04}_{-0.04}~ (40\%)}$ & $\mathrm{0.13^{+0.08}_{-0.08}}$ & $\mathrm{1 ~ (fixed)}$ & $\mathrm{0.23^{+0.04}_{-0.04}~ (54\%)}$ \\ \\
\enddata
\tablecomments{Same as \autoref{tab:regression_arnaud} but assuming each system follows the mass-dependent pressure profile from \citet{Battaglia10}.}
\end{deluxetable*}
There are a few ways to investigate the trends between X-ray and SZ properties. In the presence of considerable scatter, regressing one variable on the other does not necessarily provide the best estimate of the true relation. In such cases, orthogonal regression is commonly used to better model the data \citep[e.g.,][]{PlanckEarlyResults}. We note that orthogonal regression always yields a steeper slope compared to regressing one variable on another when there is significant scatter. Orthogonal fits were performed using the {\it{bivariate correlated errors and intrinsic scatter}} (BCES) method following the formalism presented by \citet{Akritas96}.

Aside from the statistical scatter, $\sigma_{x,y}$, it is useful to quantify the additional spread in the data known as intrinsic scatter, $\sigma_{int}$.
The total scatter, $\sigma_{tot}$, includes both of these such that
\begin{equation}
\sigma_{tot} = \sqrt{\sigma_{int}^{2} + \langle \sigma_{y}\rangle^{2} + (\beta \langle\sigma_{x}\rangle)^{2}} 
\end{equation}
We estimated $\sigma_{tot}$ from the distribution of residuals about the mean relation bounded by the $25^{th}$ and $75^{th}$ percentiles. We matched this 50\% probability to that in a normal distribution where 50\% is bounded by $\pm 0.67 \sigma$. The uncertainties on the intrinsic scatter were estimated by bootstrap re-sampling.

\begin{figure*}[t]
    \subfloat{{\includegraphics[width=0.49\textwidth]{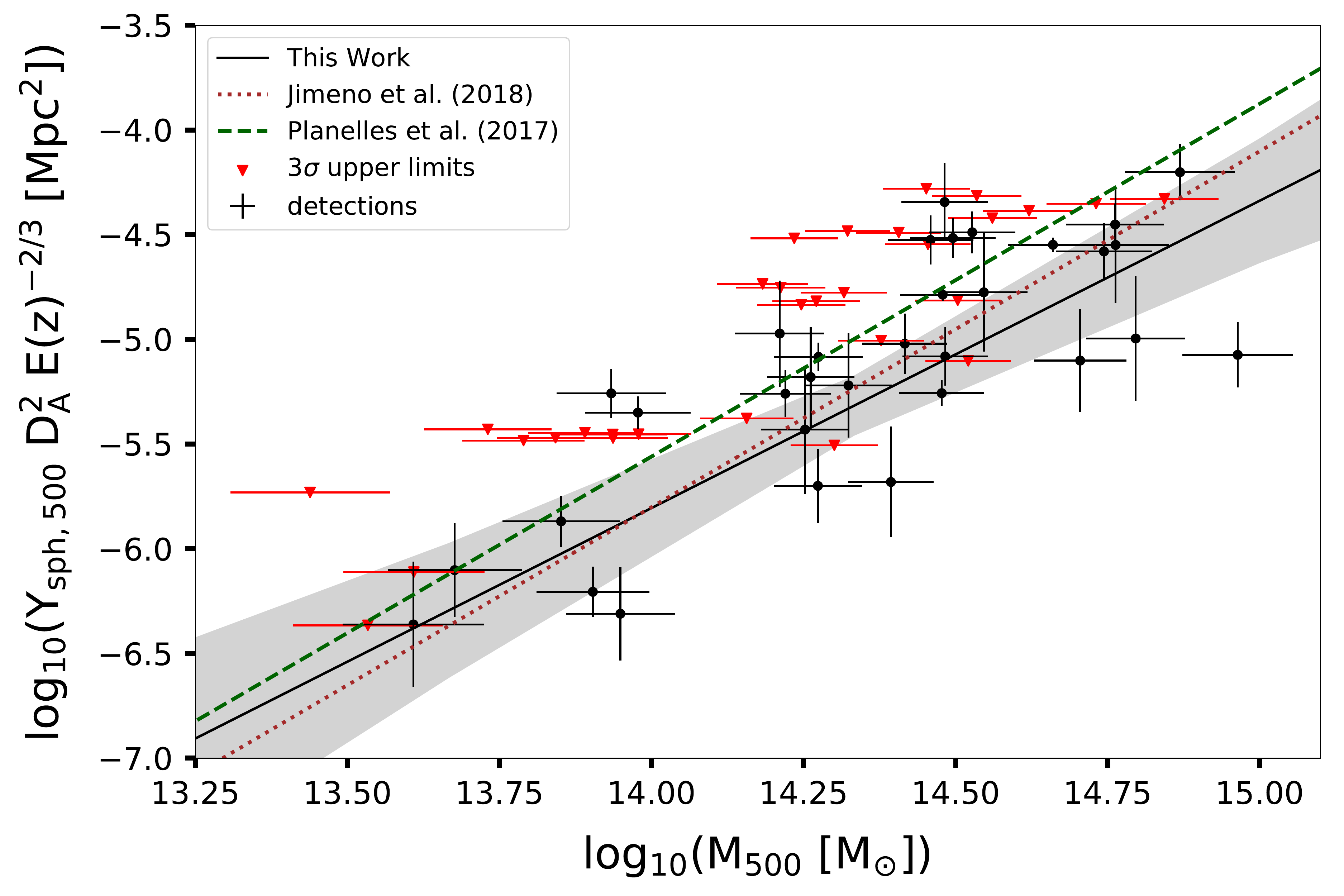} }}%
    \subfloat{{\includegraphics[width=0.49\textwidth]{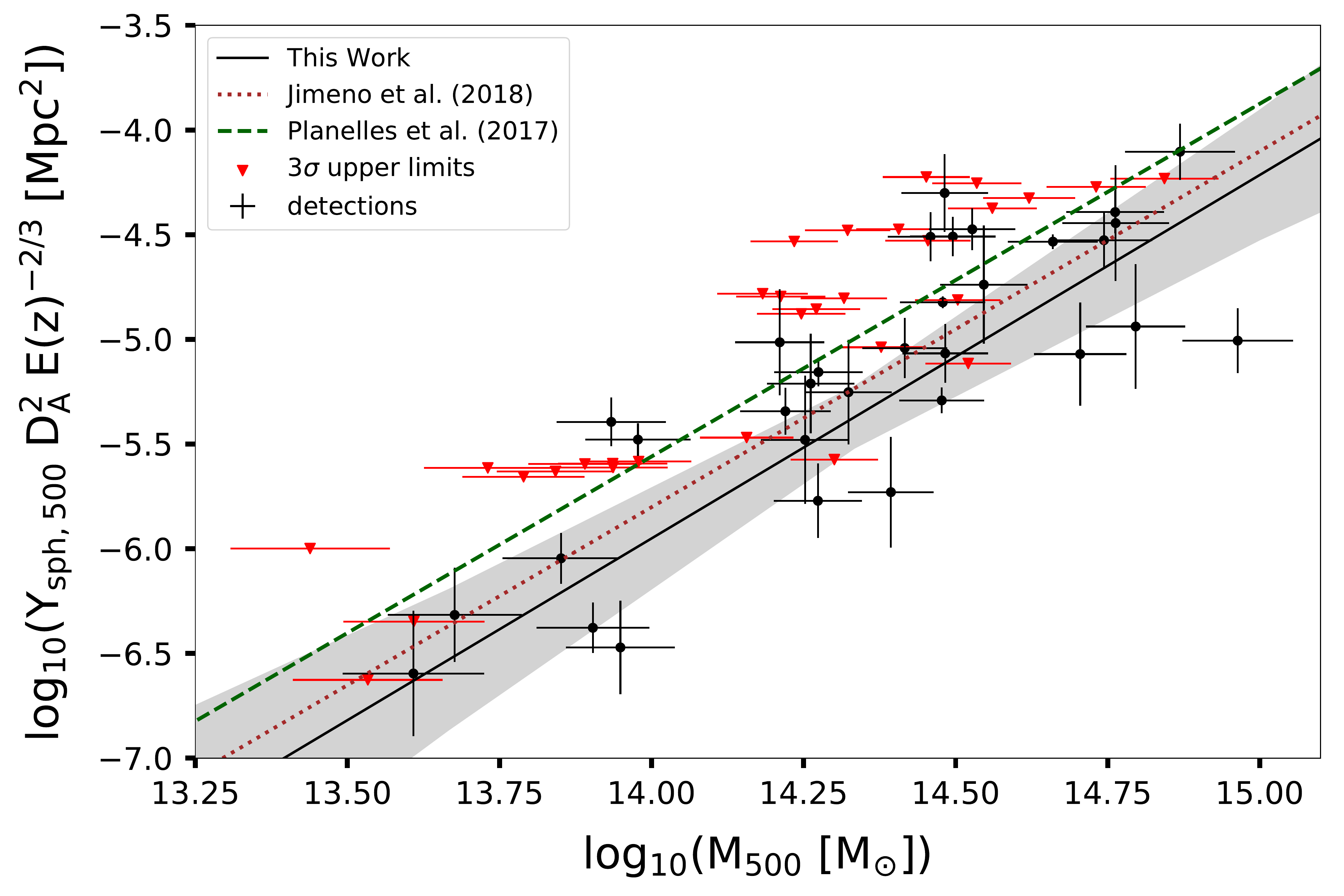} }}%
    \newline
    \subfloat{{\includegraphics[width=0.49\textwidth]{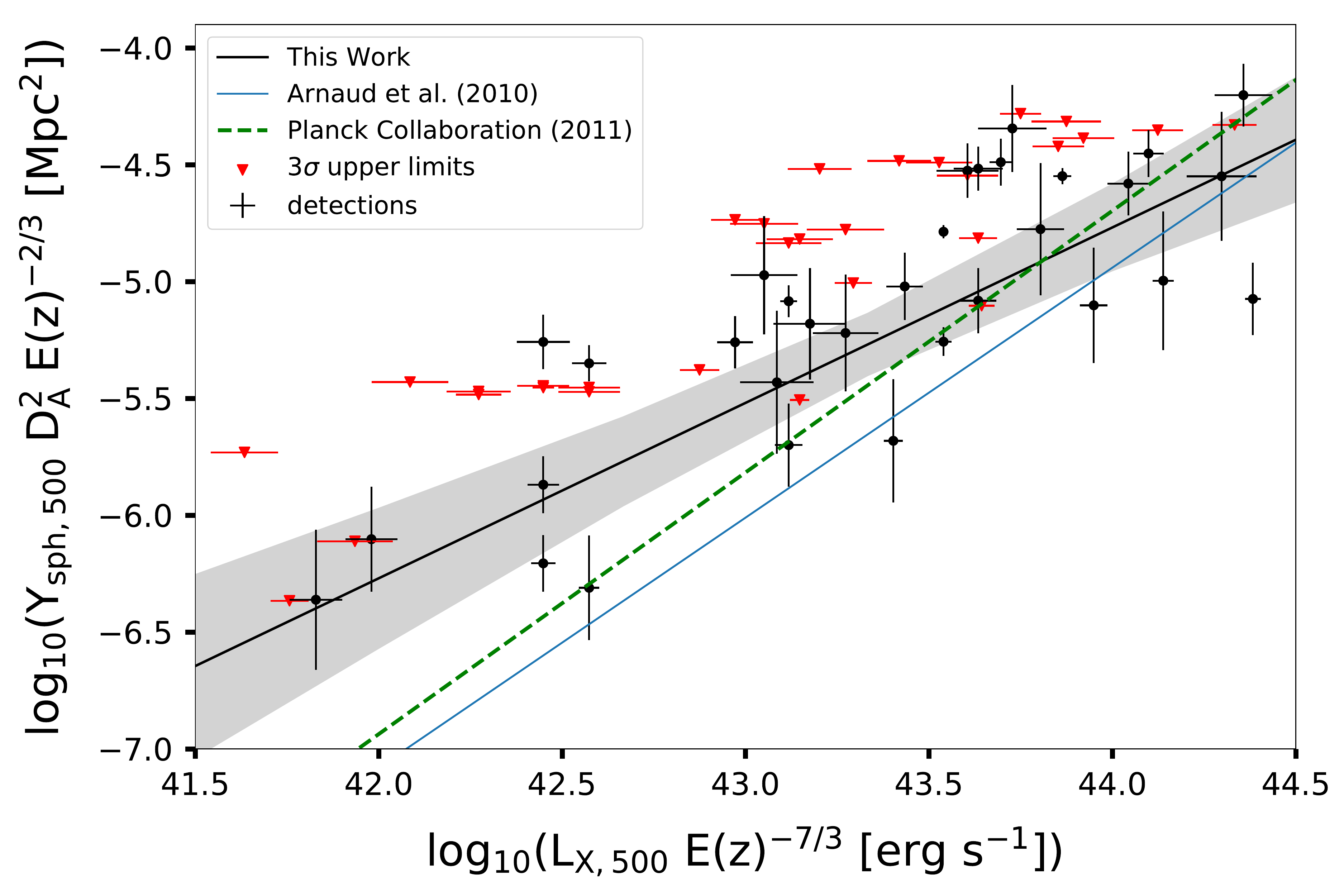} }}%
    \subfloat{{\includegraphics[width=0.49\textwidth]{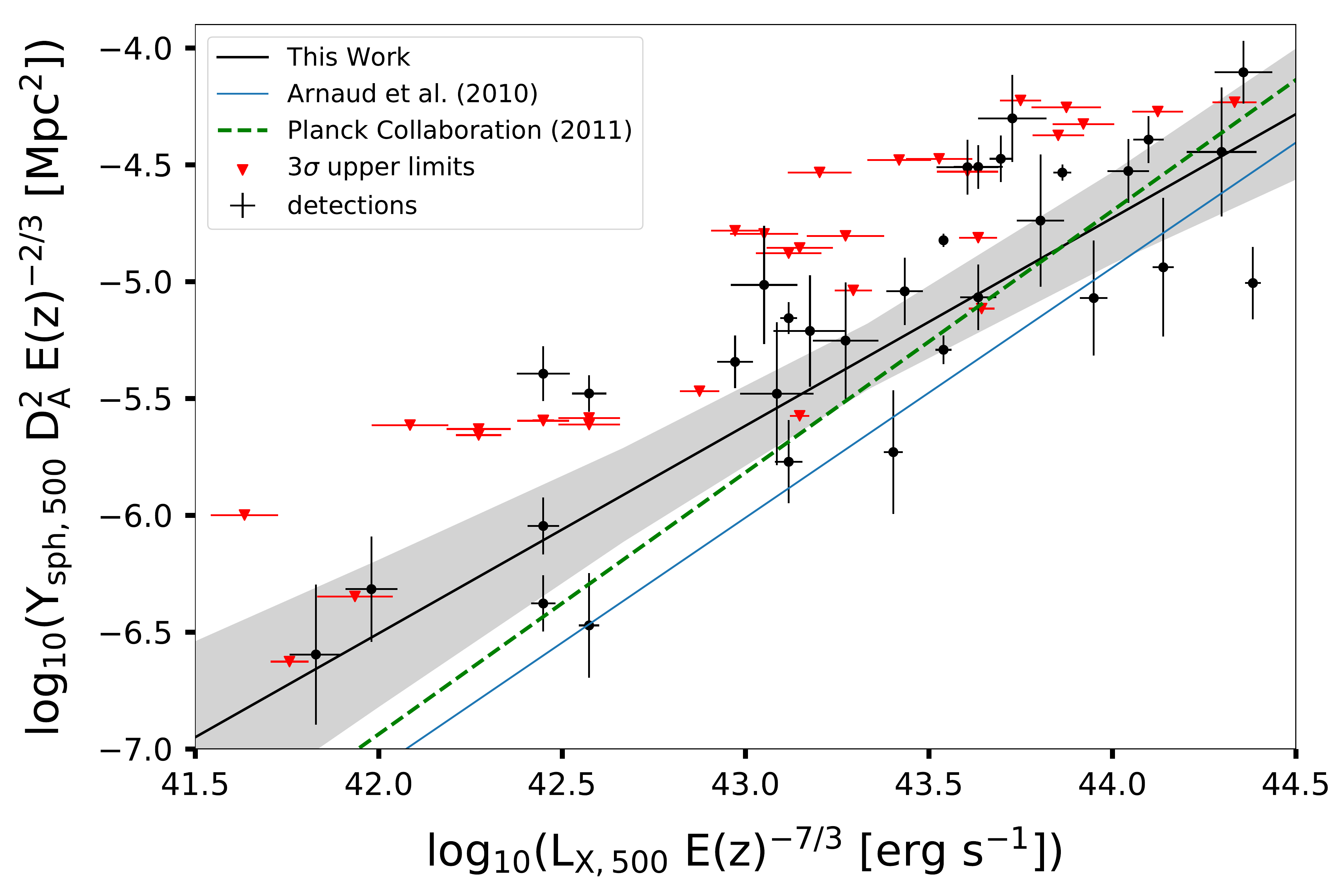} }}%
    \newline
    \subfloat{{\includegraphics[width=0.49\textwidth]{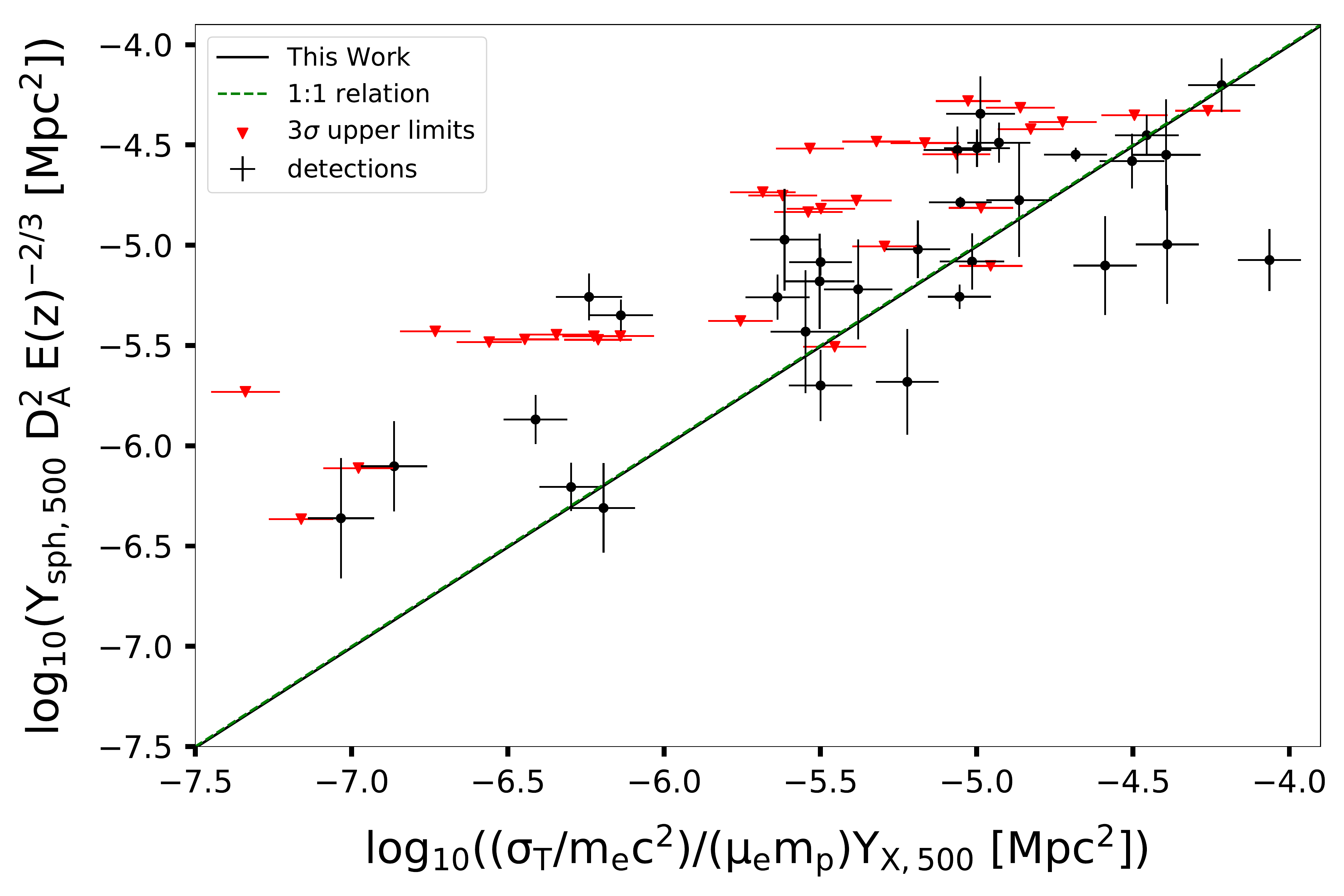} }}%
    \subfloat{{\includegraphics[width=0.49\textwidth]{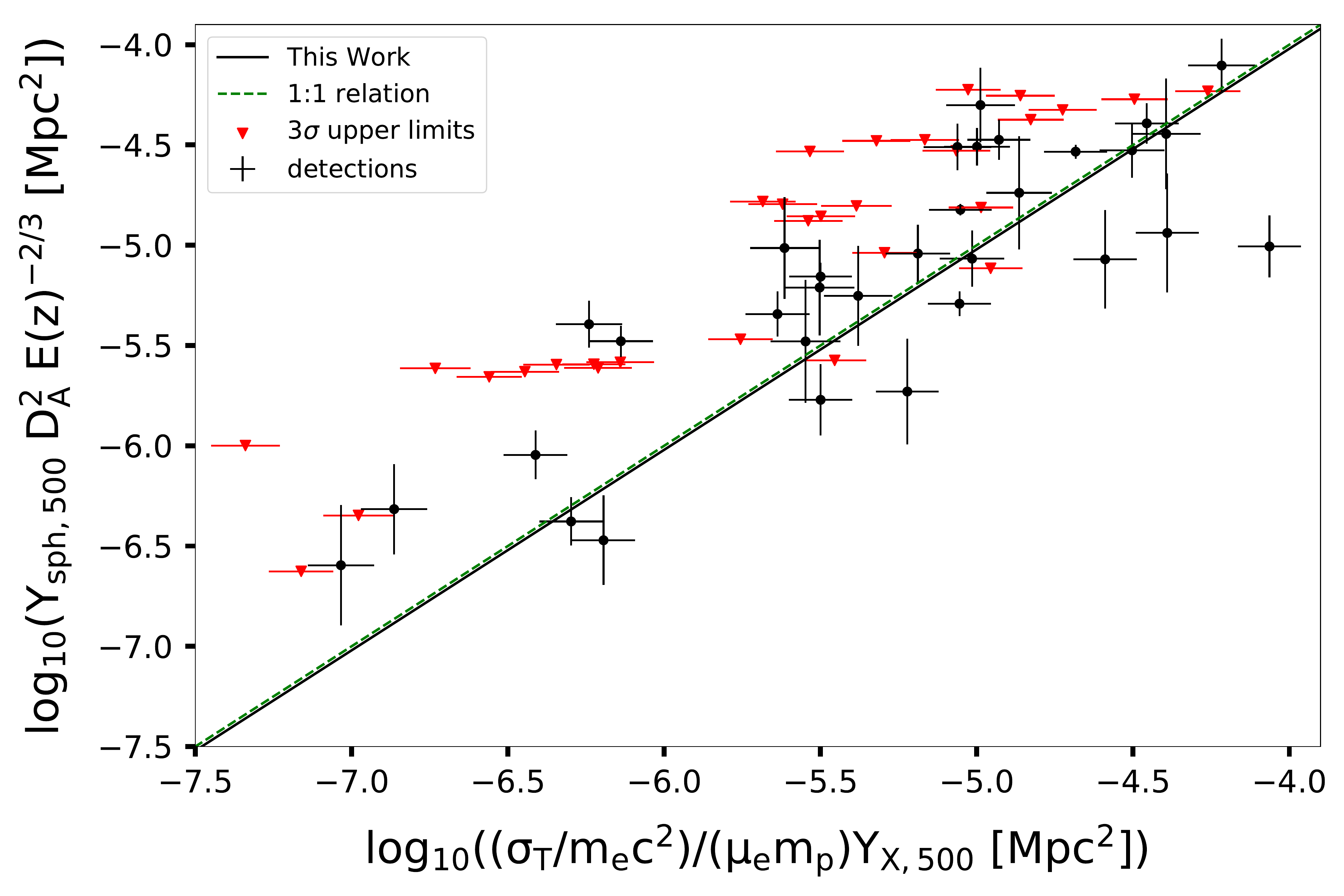} }}%
    \caption{The left column shows the SZ data using the UPP and the right column is from the BPP. The solid black lines in each plot represents the best fit using our Bayesian method on bias corrected data described in \autoref{sec:regression_methods}. The black data with error bars represent the SZ detections with SNR $>$ 1.35 while the red triangles are the nondetections placed at $3\sigma$. The shaded regions in the top two rows denote the 95\% confidence bands around the best fit relation. Results from previous studies are also shown for comparison. In the bottom row, the black line and dashed green line appear to lie on top of each other, but the black line is slightly below the dashed green line (1:1 relation).}%
    \label{fig:regression}%
\end{figure*}

We also performed linear regression using a Bayesian method, following a similar approach to \citet{Isobe86} in order to account for censored data. We maximized the likelihood function of our data, $\psi = \{x,y,\sigma_{x},\sigma_{y}\}$, given the model parameters $\theta = \{\alpha, \beta, \sigma_{int}\}$. The likelihood function can be expressed as
\begin{equation}
L(\theta|\psi) \propto \prod_{i \in D}^{m} P_{D}(\psi_{i}|\theta)\prod_{j \in C}^{n} P_{C}(\psi_{j}|\theta)
\end{equation}
\vspace{0.1cm}
\begin{equation}
    P_{D} \propto f(\psi_{i}) = \frac{\mathrm{exp}\Big(\frac{(y_{i} - \alpha - \beta x_{i})^{2}}{-2\big(\sigma_{int}^{2} + \sigma_{y,i}^{2} + (\beta \sigma_{x,i})^{2}\big)}\Big)}{\sqrt{2 \pi \big(\sigma_{int}^{2} + \sigma_{y,i}^{2} + (\beta \sigma_{x,i})^{2}\big)}}
\end{equation}
\vspace{0.1cm}
\begin{equation}
    P_{C} \approx \int_{-\infty}^{\psi_{j}}f(z)dz
\end{equation}
where $\prod_{D}^{m}P_{D}$ represents the product of conditional probabilities over $m$ detections and $\prod_{C}^{n}P_{C}$ is the same for censored data over $n$ nondetections. For censored data, $y_{i}$ becomes an upper limit on the SZ flux which was calculated at the $3\sigma$ level using the median flux uncertainty of all objects within $-1.35<SNR<1.35$. In the Bayesian approach, $\sigma_{int}$ is estimated from the posterior distribution.  

We also corrected our data for Malmquist bias. This bias arises in any flux-limited sample; it is the tendency to detect brighter objects near the flux limit as they can be seen at larger distances. Corrections for this bias has be done by \citet{Vikhlinin09} and \citet{PlanckSZCatalog} and we adopt their same method. The mean bias, {\it{b}}, is denoted by \begin{equation}
\label{eq:bias}
\mathrm{ln}~b = \langle \mathrm{ln}~ L - \mathrm{ln}~ L_{0} \rangle = \mathrm{\frac{exp(-x_{min}^2/2\sigma^{2})}{\sqrt{\pi /2}~ erfc(x_{min}/(\sigma \sqrt{2}))}~\sigma}
\end{equation}
where $L$ is the ``true luminosity'', $L_{0}$ is the measured luminosity, $x_{min} = \mathrm{ln}~f_{min} - \mathrm{ln}~f_{0}$ with $f_{min}$ being the minimum detected flux, $\sigma$ is the log-normal scatter in the relation. Since we do not know the scatter a priori, we estimated it empirically using $\sigma_{tot}$ as described above.

Malmquist bias corrections required quantifying the survey limitations. The two limits used in this study were the selection of SZ sources based on SNRs, and the cutoff in the X-ray photon count rate which was 0.002 $\mathrm{counts~s^{-1}}$ from \citet{Henry06}. The SZ flux limit was estimated by fitting a line to predict the SZ flux as a function of SNR. This line was extrapolated to the SNR cutoff (1.35), yielding a flux limit of 0.0019 arcmin$^{2}$. The survey limit in mass can be scaled from the X-ray photon count rate which is proportional to the X-ray flux and luminosity. We then used the $L_{X}-T_{X}$ relation from \citet{White97} and the $M_{500}-T_{X}$ relation from \citet{Kettula15} to scale the count rate limit to a mass limit. A similar procedure was done for the corrections on $Y_{X}$. The final numerical results, including bias corrections, are included in \autoref{tab:regression_arnaud} and \autoref{tab:regression_battaglia}.
\section{Analysis and Discussion \label{sec:analysis}}

\subsection{Comparisons of Scaling Relations \label{sec:scalings}}

Here we discuss the values of the best fit parameters for the scaling relations and their implications. In particular, we investigate the differences in our adoption of pressure profiles i.e., UPP vs. BPP. We specifically focus on the estimated parameters using our Bayesian technique and bias corrected data. For simplicity and throughout the remainder of this section, we reduce the expression $Y_{sph,500}D_{A}^{2}E(z)^{\gamma} \rightarrow Y_{SZ,500}$ where $\gamma$ is specified in \autoref{tab:regression_arnaud} and \autoref{tab:regression_battaglia} for each relation.

The slope of the $Y_{SZ,500}-M_{500}$ relation using the BPP is slightly steeper than the self-similar prediction (5/3), but it is consistent within the uncertainties. It is also in agreement with the recent observational work of \citet{Jimeno18} and the theoretical AGN feedback simulations from \citet{Planelles17}, reporting slopes of 1.70 and 1.685 respectively (shown in \autoref{fig:regression}). On the other hand, the relation found using the UPP suggests a much flatter slope of $1.32 \pm 0.21$ that is less consistent with the self-similar scaling but is still within 2$\sigma$. The intrinsic scatter is quite large ($\sim 80\%$) compared to other studies. Simulations predict a much smaller scatter of $\sim 5-15\%$ \citep{Nagai06, Planelles17}. We suggest this large intrinsic scatter is due to the lack of independent measurements of $T_{X}$ which were used to estimate the masses. We discuss this further in the next section.

Our results of the $Y_{SZ,500}-L_{X,500}$ relation suggest a slope of $0.89^{+0.09}_{-0.08}$ using the BPP. This result suggests $\sim 4\sigma$ differences from the self-similar slope of $5/4$ and about $\sim 3 \sigma$ differences from the slope of 1.07 reported by \citet{Arnaud10} (albeit their results are reported in the 0.1\textendash 2.4 keV band while ours are in the 0.5\textendash 2.0 keV band). The bias corrected slope of 1.12 found by \citet{PlanckEarlyResults} within the 0.5\textendash 2.0 keV band is within $\sim 3\sigma$ of the slope found using our Bayesian method. These authors used orthogonal regression, however, and their result is in excellent agreement with our orthogonal slope of $1.07 \pm 0.12$. We also note our data seem to systematically lie above the relation of \citet{PlanckEarlyResults} with a scatter of $\sim 80\%$ which is nearly twice their value. 

The $Y_{SZ,500}-Y_{X,500}$ relation gives insight into the amount of inhomogeneity or clumpiness in the ICM as well as the concentration of the gas toward the inner regions \citep{Planelles17}. Since $Y_{X}$ is the analog to the SZ luminosity, one might expect the $Y_{SZ,500}/Y_{X,500}$ ratio to be unity, however, $Y_{SZ,500}/Y_{X,500}$ has been found to be slightly smaller than unity from other observational work \citep[e.g.,][]{Arnaud10,PlanckEarlyResults,Rozo12}. This ratio is really a comparison of the mass weighted temperature to the spectroscopic temperature and is expected to be smaller than unity for a decreasing temperature profiles \citep{Arnaud10}. Our results suggest a ratio of 0.98 and 0.95 using the UPP and BPP respectively, which is in agreement with previous studies. We also point out that objects with low $Y_{X,500}$ values tend to systematically lie above the line of unity for both pressure profiles. This may be an important result arising from the assumed pressure profiles. We discuss this further in the next section.
\subsection{Caveats}
Here we analyze the limitations of our data and the validity of the necessary assumptions made. We also discuss subsequent steps to improve future joint X-ray\textendash SZ studies.

One limitation in this study was that the only raw X-ray measurements available were the count rates within an aperture of $R_{200}$. The count rates were converted into luminosities which were used to estimate the temperatures using the $L_{X}-T_{X}$ relation from \citet{White97}. We then used those temperatures to estimate $M_{500}$ using the $M_{500}-T_{X}$ relation from \citet{Kettula15}. Ultimately, the values of $T_{X}$ and $M_{500}$ were derived from the count rates.
\startlongtable
\begin{deluxetable*}{ccrccccl}
\tablecaption{Potential Cluster Candidates
\label{tab:new_objects}}
\tablewidth{0.5\textwidth}
\tablehead{
\colhead{RA} & \colhead{Dec} & \colhead{SNR} & \colhead{$\mathrm{Y_{SZ}}$} & \colhead{ID} & $\alpha \mathrm{(J2000)}$ & $\delta \mathrm{(J2000)}$ & \colhead{z} \\
& \colhead{} & \colhead{} & [$10^{-4}$ arcmin$^{2}$] & \colhead{} & \colhead{} & \colhead{} & \colhead{}}
\colnumbers
\startdata
18:35:40.8 & 66:47:05.9 & 4.07 & $6.2 \pm 1.5$ & \nodata & \nodata & \nodata & \nodata\\
\hline
17:46:36.9 & 68:44:59.9 & 4.09 & $6.9 \pm 1.7$ &  WARP J1746.3+6849W & 17:46:19.0 & +68:49:50.0 & 0.203 \\
& & & & 400d J1746+6848 & 17:46:29.1 & +68:48:54.0 & 0.217 \\
& & & & WARP J1746.3+6849E & 17:46:33.0 & +68:48:50.0 & 0.307 \\
\hline
17:48:26.4 & 64:59:07.1 & 4.22 & $6.2 \pm 1.5$ & WHL J174744.2+645225$^{~a}$ & 17:47:44.2 & +64:52:25.0 & 0.3758  \\
\hline
17:47:45.6 & 64:52:14.0 & 4.67 & $6.6 \pm 1.4$ & WHL J174744.2+645225$^{~a}$ & 17:47:44.2 & +64:52:25.0 & 0.3758 \\
\hline
17:16:41.6 & 67:10:41.1 & 4.83 & $6.9 \pm 1.4$ & RX J1716.4+6708 & 17:16:49.6 & +67:08:30.0 & 0.813 \\
\hline
18:44:21.9 & 64:17:34.2 & 5.23 & $6.6 \pm 1.3$ & WHL J184505.8+642618 & 18:45:05.8 & +64:26:18.1 & 0.3002 \\
& & & & WHL J184502.2+641754 & 18:45:02.2 & +64:17:54.3 & 0.4353\\
\hline
17:55:00.7 & 64:21:15.7 & 5.50 & $6.4 \pm 1.2 $ & WHL J175517.4+641630$^{~b}$ & 17:55:17.4 & +64:16:29.0 & 0.2837 \\
\hline
18:24:33.4 & 69:20:02.9 & 6.77 & $8.8 \pm 1.3$ & \nodata & \nodata & \nodata & \nodata \\
\hline
18:21:24.0 & 64:21:35.3 & 9.45 & $20.0 \pm 2.1$ & WHL J182203.5+642301 & 18:22:03.5 & +64:23:01.1 & 0.3315\\
\hline
18:31:27.7 & 62:18:57.0 & 13.38 & $15.9 \pm 1.2$ & PSZ1 G091.82+26.11 & 18:31:08.2 & +62:14:51.7 & 0.24 \\
\enddata
\tablenotetext{$a$}{These are overlapping regions that contain the same cluster within $10^{\prime}$.}
\tablenotetext{$b$}{This region contains a confirmed cluster detected by \citet{Gioia03} which was the optical identification program used to construct the X-ray NEP catalog. \citet{Henry06} did not include this in their catalog as it did not meet all requisites for their final catalog.}
\tablecomments{Presented are the coordinates of the regions which have high SNR values (SNR $\geq 4$) and the number of groups/clusters within a $10^{\prime}$ radius according NED.}
\end{deluxetable*}
The measurements of total X-ray luminosities are probably a strong source for the large amounts of intrinsic scatter in our scaling relations. X-ray scaling relations involving the total luminosity show the largest amounts of scatter, especially when the cores have not been excised. Virialized systems may eventually develop cores with very short cooling times that boost their X-ray luminosities. Due to this excess luminosity, the cores ($R \lesssim 0.15~R_{500}$) are typically ignored in scaling relation studies \citep[e.g.,][]{Mantz16}. The X-ray luminosities used in this study were calculated without excising the central regions, so we expect at least a few cool cores to be present in our sample. An example of a potential cool core system is the object containing the highest mass, luminosity, and $Y_{X,500}$ in \autoref{fig:regression}. This datum seems to lie significantly below the predicted line for all three scaling relations. The presence of a strong cool core would cause us to overestimate its mass and shift the datum to the right.

Systems may also be morphologically disturbed which tend to down scatter the luminosity from the mean relation \citep{Pratt09}. The fact that different systems exhibit a variety of morphologies can contribute to large observed scatter \citep{Marrone12}. It has been shown that the scatter in X-ray scaling relations is reduced once the cores have been excised and when morphologically disturbed systems have been accounted for \citep{Pratt09,Arnaud10}.

The unresolved nature of our SZ measurements required making assumptions about the pressure profiles for each system. We first assumed each system followed the UPP from \citet{Arnaud10}. While this may be applicable to massive clusters, it is probably a poor assumption to make for groups if there is significant AGN feedback. Simulations and observations have shown that the UPP can overestimate $Y_{SZ,500}$ by almost an order of magnitude toward masses $M_{500} \sim 10^{13} M_{\odot}$ \citep{LeBrun15,Lim18}. This may be causing our objects with low $Y_{X,500}$ values to systematically lie above the 1:1 line in the bottom-left panel of \autoref{fig:regression}. To account for this, we also considered the mass- and redshift-dependent pressure profiles derived by \citet{Battaglia10}. The BPP can lower $Y_{SZ,500}$ by $\lesssim 30\%$ relative to the UPP for systems $M_{500} \lesssim 10^{14.5} M_{\odot}$. 

While the BPP may be a more realistic for groups compared to the UPP, it does not evade the inevitable problem of assuming a pressure profile. In fact, the choice between the UPP and BPP had significant effects on the derived scaling relations. Specifically, we found the UPP increases the intercept by $\sim 3\sigma$ and $\sim 4\sigma$ and decreases the slope by $\sim 2\sigma$ and $\sim 1\sigma$ compared to the BPP for the $Y_{SZ,500}-M_{500}$ and $Y_{SZ,500}-L_{X,500}$ relations respectively. This is not confounding as one would expect each system to experience different amounts of AGN feedback while also existing in a variety of virialization states. Again, these differences are expected to be more pronounced in groups than clusters. 

Our last noteworthy assumption was made in calculating $M_{gas,500}$. After solving for the emission measure using \autoref{eq:APEC}, we adopted an isothermal $\beta$ model for the density distribution, assuming all systems follow a density distribution with $\beta = 2/3$. The $\beta$ model is a good first order approximation, however, others have argued for better models of the density distribution \citep[e.g.,][]{Vikhlinin06,Hallman07}. Even with the assumption of the isothermal $\beta$ model, it has been shown that groups typically exhibit smaller values of beta (i.e., flatter profiles) which would increase our estimates of the total gas mass in lower mass systems. In order to acquire more accurate measurements of $Y_{X,500}$, it would be ideal to use systems that are resolved in X-rays.

\subsection{Future Work}
In this study, we quantified the effects of assuming different pressure profiles to calculate $Y_{SZ,500}$. The large amounts of intrinsic scatter compared to other studies suggests that scaling relations can be improved with resolved data. In order to avoid making strong assumptions, resolved SZ or X-ray surface brightness profiles are needed to construct accurate pressure profiles for individual halos. X-rays can provide temperature and density profiles near the central regions, which is useful for constraining the amount of the SZ flux stemming from the outer regions ($\gtrsim R_{500}$). We can then extrapolate the profiles to large radii to get a sense of how effective AGN are at driving gas outward. Moreover, one could directly measure the pressure profiles at large radii using resolved SZ data. This has been done using nearby, massive clusters \citep[e.g.,][]{PlanckVirgo}, however, little attention has been given to groups since they produce weaker signals. Resolved observations for these low-mass systems would provide the best evidence for the effects of AGN feedback in groups.
\begin{figure*}[!t]
    \subfloat{{\includegraphics[width=0.45\textwidth]{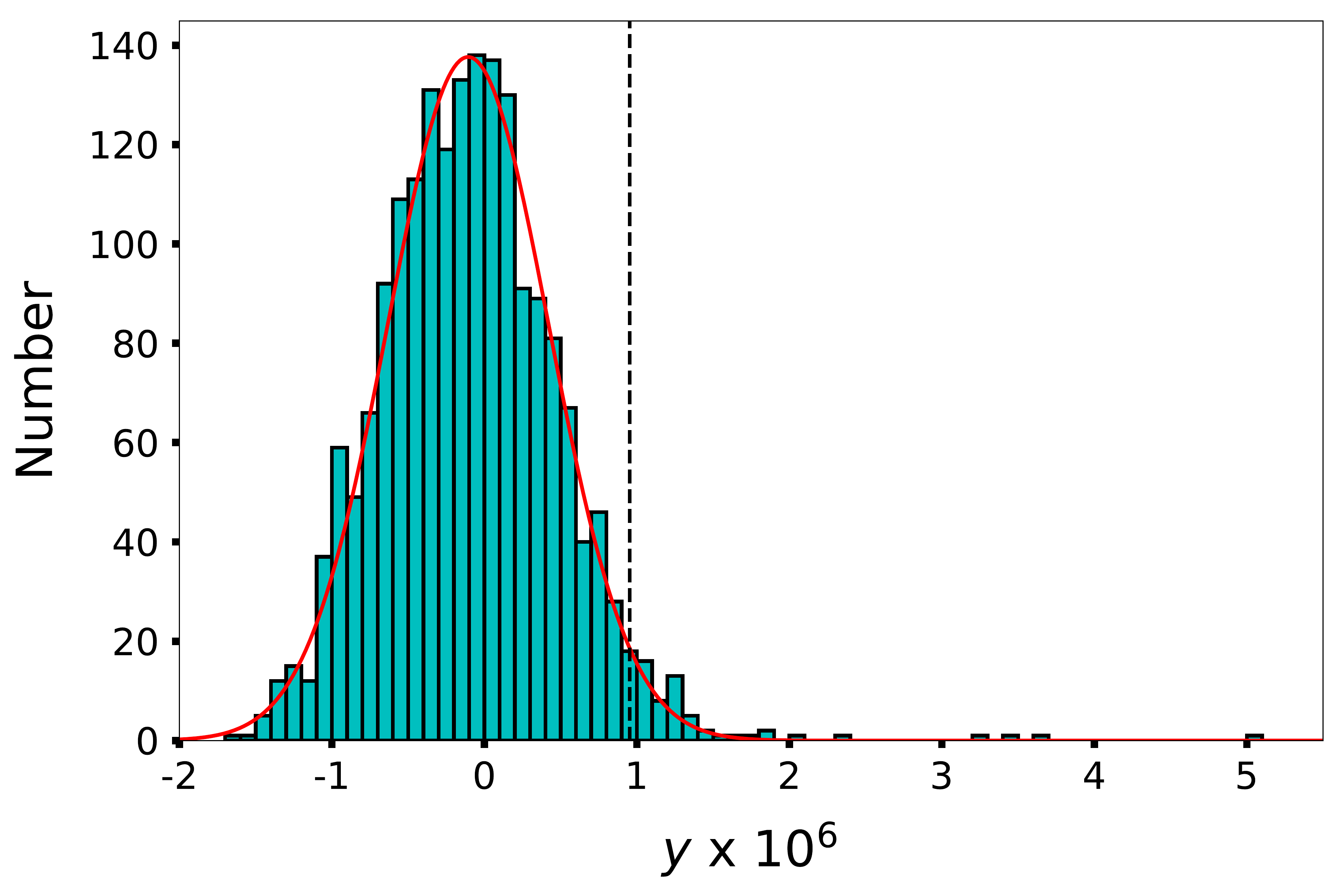} }}%
    \subfloat{{\includegraphics[width=0.45\textwidth]{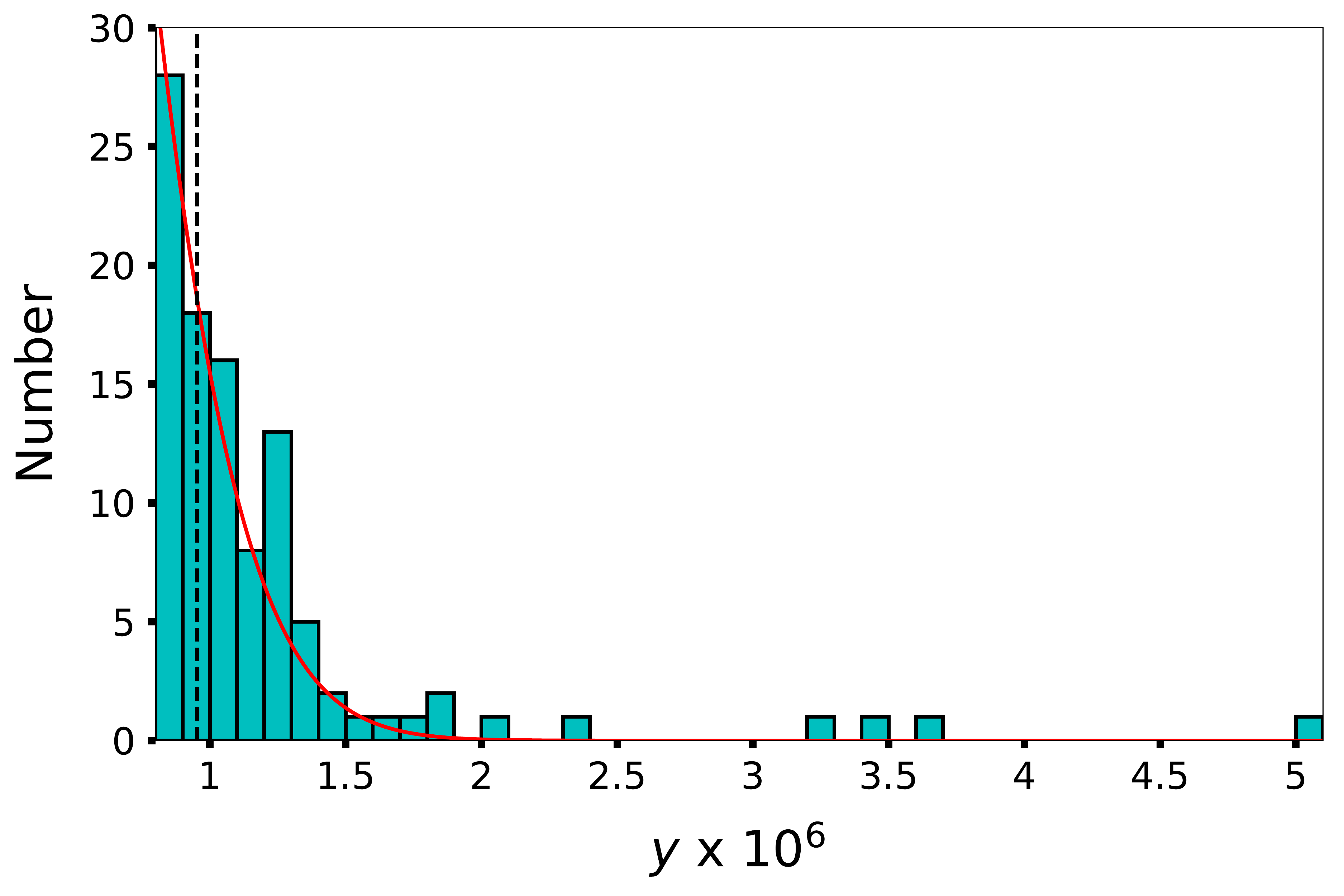} }}%
    \caption{Histogram showing the results of a blind search for strong SZ signals near the NEP. The left panel shows the full distribution of the average values of circles for the entire random array. The right panel is zoomed in to clearly show the outliers. In both panels, the red line represents the best fit Gaussian curve and the dashed gray line indicates $2\sigma$ above the mean.}%
    \label{fig:circles}%
\end{figure*}
In addition to the X-ray detected groups/clusters near the NEP, we searched the y-maps for any objects that may have been missed by the NEP X-ray survey. Due to the redshift independent nature of the SZ effect and the flattening of the angular distance from $0.5 \lesssim z \lesssim 2$, it is possible to detect groups/clusters in the $y$-maps but not in X-rays. We started a search for these objects using an array of $10^{\prime}$ circles across the NEP. There were a total of 1874 circles inside the limits of the NEP study while containing $\geq 75\%$ of uncontaminated area. The distribution of the average values of these circles is presented in \autoref{fig:circles}.

We further investigated the 59 circles with average values lying $2\sigma$ above the mean. These circles were reduced to 31 group/cluster candidates. Since we did not have X-ray data for these objects \textemdash meaning we did not have estimates of $M_{500}$ \textemdash we fit PSF profiles to them in order to get a SNR. We found 30 candidates with $SNR > 1.35$ and 10 with $SNR > 4$. Using the 10 highest SNR regions, we searched the NASA Extragalactic Database for archival data on these candidates within a $10^{\prime}$ radius. We found that $8/10$ of the candidates were confirmed clusters that went undetected in the NEP X-ray catalog. There was one exception of a cluster that was excluded from the NEP X-ray catalog that was previously detected by the NEP optical program \citep{Gioia03}. The results are presented in \autoref{tab:new_objects}.  While the NEP X-ray catalog is one of the deepest X-ray catalogs to exist, these results show there are many groups/clusters in this region that need follow-up X-ray observations.

\section{Summary and Conclusions \label{sec:conclusion}}
The main goal of this work was to measure the SZ\textendash X-ray scaling relations of galaxy groups and clusters near the NEP. The NEP is perhaps the best region of the sky with both deep X-ray and SZ data, allowing one to investigate a considerable sample of individual galaxy groups and low-mass clusters. We highlight the main aspects and conclusions of this work below.
\begin{enumerate}
    \item This study investigated a sample of 62 galaxy groups and clusters detected in X-rays by \citet{Henry06} and optically by \citet{Gioia03}. We were able to extract the SZ signal from 32 unresolved systems with SNRs $>$ 1.35. The remaining nondetections were placed as upper limits at the 3$\sigma$ level when performing linear regression.
    \item The unresolved nature of the X-ray and SZ data required making assumptions about the pressure profile of each system in order to convert the total SZ flux into the fraction inside \Rvir. We considered two pressure profiles: the UPP, which is the observational ``universal pressure profile'' of galaxy clusters from \citet{Arnaud10}, and the BPP, which is the mass-dependent pressure profile derived from the simulations of \citet{Battaglia10}.
    \item The scaling relations were derived using a Bayesian linear regression technique that takes into account nondetections. We estimated a slope of $1.73^{+0.19}_{-0.18}$ for the $Y_{SZ,500}-M_{500}$ relation using the BPP, which is consistent with the self-similar prediction as well as other observational and theoretical studies of massive clusters. The derived $Y_{SZ,500}-L_{X,500}$ scaling relation using the BPP yielded a slope of $0.89^{+0.09}_{-0.08}$, which is much flatter than the self-similar slope as well as those reported by other observational works; we note, however, that our orthogonal regression produces a steeper slope of $1.07 \pm 0.12$ that is consistent with previous studies. The $Y_{SZ,500}-Y_{X,500}$ relation using the BPP data yields an intercept value of $-0.02^{+0.07}{-0.08}$ when the slope was fixed to unity; this is slightly smaller than zero which is consistent with previous findings.
    \item The choice between the UPP and the BPP has significant effects on the scaling relations. The derived scaling relations using the UPP yields larger intercepts and flatter slopes compared to the BPP. We considered the BPP to be a more realistic description of the pressure profiles across our mass spectrum, however, they need to be confirmed with resolved X-ray and/or SZ observations toward the low-mass regime.
    \item We conducted a blind search for strong SZ signals near the NEP aside from the groups and clusters cataloged by \citet{Henry06}. The search yielded 10 regions with strong SZ signals above the background, and 8/10 candidates have known galaxy clusters in the vicinity after searching NED. These are excellent targets for follow-up X-ray observations that would help build more complete sample of galaxy groups and clusters near the NEP.
\end{enumerate}
This study demonstrates one can conduct a sizable study of galaxy groups and low-mass clusters with current X-ray and SZ data. Given the low sensitivity of the $Planck$ y-maps, having available X-ray data allows one to exploit low-SNR SZ signals. In addition, we stress the importance of using resolved X-ray data if one wants to conduct an accurate study of the SZ scaling relations using the current y-maps from \citet{PlanckYMAP}.

We thank the many individuals who offered guidance and insight, including Zhijie Qu, Chris Miller, Yanson Yun, and the anonymous referee, whose comments and suggestions helped improve this work. We are grateful for support from NASA through the Astrophysics Data Analysis Program, awards NNX15AM93G and 80NSSC19K1013. The work is based on observations obtained with \textit{Planck}, http://www.esa.int/Planck, an ESA science mission with instruments and contributions directly funded by ESA Member States, NASA, and Canada. This research made use of the High Energy Astrophysics Science Archive Research Center (HEASARC) and the NASA/IPAC Extragalactic Database (NED), which are supported by NASA.
\newpage
\bibliography{main.bib}



\end{document}